\documentclass[aps,twocolumn,showpacs,showkeys,amsmath,amssymb,superscriptaddress,nofootinbib,floatfix,longbibliography]{revtex4-1}

\usepackage[utf8]{inputenc}
\usepackage{epsfig,bm,slashed}

\begin{document}
\title{Off-shellness in generalized parton distributions and form factors of the pion}

\author{Vanamali Shastry}
\email{vanamalishastry@gmail.com}
\thanks{\\Present address: Center for  Exploration  of  Energy  and  Matter, Indiana  University, Bloomington,  IN  47403,  USA.}
\affiliation{Institute of Physics, Jan Kochanowski University, 25-406 Kielce, Poland}

\author{Wojciech Broniowski}
\email{Wojciech.Broniowski@ifj.edu.pl}
\affiliation{H. Niewodnicza\'nski Institute of Nuclear Physics PAN, 31-342 Cracow, Poland}
\affiliation{Institute of Physics, Jan Kochanowski University, 25-406 Kielce, Poland}

\author{Enrique Ruiz Arriola}
\email{earriola@ugr.es}
\affiliation{Departamento de F\'{\i}sica At\'{o}mica, Molecular y Nuclear and Instituto Carlos I de  F{\'\i}sica Te\'orica y Computacional, 
Universidad de Granada, E-18071 Granada, Spain}

\date{\today}  

\begin{abstract}

We study the effects of off-shellness in the generalized parton distributions of the pion. On general grounds, these distributions exhibit a richer structure 
than in the on-shell case due to absence of the crossing symmetry. In particular, their moments involve additional terms odd in the skewness parameter, associated with new form factors. We bring up relations between the off-shell charge and gravitational form factors, as well as the pion form factor, and discuss their derivations based on the Ward-Takahashi identities. We illustrate the features at the (leading-$N_c$) one-quark-loop level with the help of the spectral quark model of the pion, constructed to embed the vector meson dominance. Simple analytic expressions for the form factors and the distributions follow. 
Thus obtained off-shell generalized parton distributions are evolved from the quark model scale to higher scales with the LO DGLAP equations.
We evaluate the corresponding Compton amplitudes which enter the cross-section for the electroproduction of the pion off the proton (the Sullivan process). 
It is found in our model that the effects of off-shellness in the generalized parton distribution are substantial, however, 
they can be largely canceled by the corresponding off-shell corrections to the pion propagator.
In particular, this is the case of the Compton form factors entering the deeply virtual Compton scattering amplitude. 
As a result, we expect small off-shellness effects in electroproduction reactions, such as the Sullivan process.
\end{abstract}

\maketitle

\section{Introduction}
This paper extends our recent study~\cite{Broniowski:2022iip} 
of off-shell effects in generalized parton distributions (GPDs) of the pion.

Exploring the non-perturbative structure of the pion and other
pseudoscalar states has been of continued interest over the recent
years~\cite{Amoroso:2022eow}.  However, the unstable nature of the
pion makes it difficult to investigate it experimentally. In
particular, the study of the features of GPDs is not possible in the
conventional deeply virtual Compton scattering (DVCS) experiments and
one has to resort to indirect methods such as the Sullivan
process~\cite{Sullivan:1971kd}. This reaction, shown in
Fig.~\ref{fig:sullifd}, involves the electroproduction of the pion off
a proton,
\begin{align}
    \gamma^* + p \to \pi^+ + n
\end{align}
(the leptonic component has been omitted). The process involves the DVCS amplitude, albeit with one of the pions off the mass shell, {\it i.e.,}
\begin{align}
    \gamma^* + \pi^{+\, *} \to \gamma + \pi^+,
\end{align}
which calls for a detailed exploration of the off-shellness effects in modeling this process.

Actually, one of the important goals of the upcoming Electron-Ion Collider (EIC) 
facility is to study the Sullivan process~\cite{Aguilar:2019teb,Arrington:2021biu,Abir:2023fpo}, while
the feasibility of extracting the pion GPDs was studied in~\cite{Chavez:2021koz,Chavez:2021llq,Chavez:2023hbl}, where also 
the beam spin asymmetry was found to be influenced by the corresponding gluon distributions. 
For further expositions into this problem see also~\cite{Qin:2017lcd,Wang:2023thl,Goloskokov:2022rtb,Goloskokov:2022mdn} and references therein.

In the past, the parton distribution functions (PDFs), which are limiting cases of the GPDs, have been studied for the pion in various experiments, such as the pion-nucleus scattering at the CERN NA3~\cite{NA3:1983ejh}, the Fermilab E615~\cite{Conway:1989fs}, or
the electroproduction at HERA~\cite{ZEUS:2002gig,H1:2010hym} (cf. a recent study of PDFs in~\cite{Barry:2021osv}).

An alternate framework is to study the GPDs on the lattice. To this end, the quasi-distributions introduced by Ji~\cite{Ji:2013dva} provide a necessary
tool~\cite{Chen:2016utp,Alexandrou:2015rja,Alexandrou:2017dzj,Zhang:2018diq,Izubuchi:2019lyk,Lin:2020ssv,Gao:2020ito,Chen:2020arf,Braun:2020ymy,Ji:2015qla,Liu:2019urm,Fan:2018dxu,Zhang:2020rsx}, as well as the related pseudo-distributions~\cite{Radyushkin:2016hsy,Radyushkin:2017cyf,Orginos:2017kos,Monahan:2016bvm,%
Monahan:2017oof,Radyushkin:2018nbf,Karpie:2018zaz,Joo:2019bzr,Joo:2019jct,Karpie:2019eiq,DelDebbio:2020rgv,Joo:2020spy,Bhat:2020ktg,Radyushkin:2019owq,Balitsky:2021qsr,Balitsky:2021cwr}.\par
Such lattice simulations have produced interesting insights into the structure of the pion. The first evaluation of the leading-twist GPD of the pion with zero skewness were reported in~\cite{Chen:2019lcm,Karthik:2021sbj}. The Large Momentum Effective Theory (LaMET) based lattice study~\cite{Chen:2019lcm} and the numerical estimations using a combination of lattice data and Bayesian statistics~\cite{Karthik:2021sbj} were used to extract the moments of the pion parton distribution functions as well as the zero-skewness GPDs. The lower moments of the pion structure functions were found to be in good agreement with the experimental data~\cite{Martinelli:1987zd,Morelli:1991gb,Best:1997qp,Detmold:2003tm}. 
Newer formalisms of extracting the GPDs of hadrons on the lattice have been proposed and are being actively implemented~\cite{Bhattacharya:2022aob,Constantinou:2022fqt,Cichy:2023dgk,Bhattacharya:2023nmv}. Detailed reviews on the status of the problem can be found e.g. in~\cite{Lin:2017snn,Cichy:2018mum,Monahan:2018euv,Zhao:2018fyu,Constantinou:2020hdm}.

From the theoretical side, various models have been applied to study the non-perturbative structure of the pion, and in particular the partonic distributions. 
Chiral symmetry is one of the prerequisites to construct reliable models for the pion, which is a pseudo-Goldstone boson of the spontaneously broken chiral symmetry. The PDFs of the pseudoscalar mesons were estimated in the Nambu--Jona-Lasinio (NJL) model (for a review, see~\cite{RuizArriola:2002wr}) in~\cite{Davidson:1994uv,Davidson:2001cc}, in the instanton liquid model in~\cite{Dorokhov:2000gu,Anikin:2000rq,Noguera:2005cc,Nam:2010pt}, and in the rainbow-diagram Dyson-Schwinger approach \cite{Weiss:1994ke,Nguyen:2011jy,Chang:2014lva}. The GPDs in NJL were explored in~\cite{Broniowski:2007si,Courtoy:2010qn}. The quasi-parton distributions were computed in NJL in~\cite{Broniowski:2017wbr,Broniowski:2017gfp,Broniowski:2017zqz} and in the instanton liquid model in~\cite{Kock:2020frx}.\par

In the present paper, we build up on our previous work~\cite{Broniowski:2022iip} and elaborate important aspects. 
In Section~\ref{sec:formal} we
present the definitions of the GPDs and establish our notation. In Section~\ref{sec:ff} we derive the Ward-Takahashi identities (WTIs) for off-shell vector and gravitational form factors and discuss the effects of off-shellness and the resulting lack of the crossing symmetry on the polynomiality of the GPDs, as well as the emergence of additional off-shell form factors. In the second part we illustrate the general results in a simple chiral quark model. In Section~\ref{sec:model}  we describe the spectral quark model (SQM) used to quantify the results and present the obtained GPDs, which are evolved to experimental scales with the DGLAP-ERBL equations. The model vector and gravitational form factors are discussed in Section~\ref{sec:vgff}. The Compton form factors obtained form our GPDs are presented in Section~\ref{sec:cff}.  
In Section~\ref{sec:offpi}, we provide a detailed discussion of off-shellness issues in hadronic processes, including the important effect of cancellation from the off-shell 
pion propagator. 

\section{Formalism}\label{sec:formal}

\begin{figure*}[tb]
    \centering
    \includegraphics[width=0.32\textwidth]{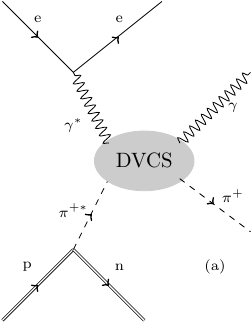} \hspace{1cm} %\\\vspace{0.2in}
    \includegraphics[width=0.32\textwidth]{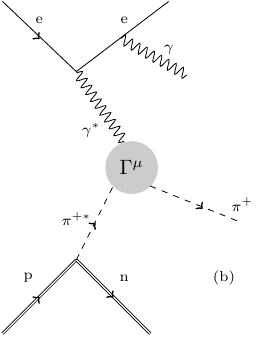}
    \caption{Diagrams for the pion electroproduction off the proton (the Sullivan process). The DVCS on the pion enters diagram (a), which interferes with the Bethe-Heitler process of diagram (b) (the corresponding diagram with the photon emission from the initial electron is not shown).
    \label{fig:sullifd}}
\end{figure*}

\begin{figure}[tb]
    \centering
    \includegraphics[width=0.355\textwidth]{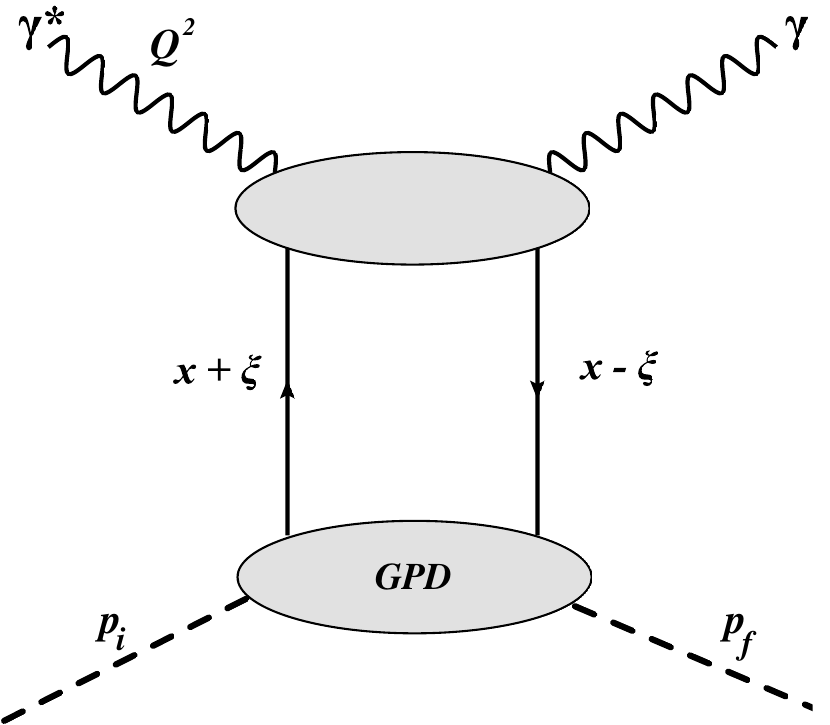}    
    \caption{DVCS amplitude of the pion decomposed into GPD and the hard perturbative kernel. The longitudinal momentum fraction of the struck parton is changed from  $x+\xi$ into $x-\xi$, where $\xi$ is the skewness parameter. \label{fig:gpd}}
\end{figure}

The Sullivan electroproduction process is depicted in Fig.~\ref{fig:sullifd}. It involves the DVCS amplitude (panel a), interfering with the Bethe-Heitler amplitude (panel b). In this reaction, the virtual pion in the DVCS or the Bethe-Heitler amplitudes is generally not on shell, hence it becomes necessary to study in detail the effects of the pion off-shellness. 
The DVCS amplitude can be factorized into a soft matrix element, corresponding to GPD, 
and a hard kernel calculable perturbatively in QCD (cf. Fig.~\ref{fig:gpd}). 

The kinematics in the
assumed symmetric notation is as follows:
\begin{eqnarray}
P^\mu=\tfrac{1}{2} (p_f^\mu + p_i^\mu), \;\; q^\mu=p_f^\mu - p_i^\mu,  \;\; \xi=-\frac{q^+}{2P^+}, \;\;  t=q^2, \nonumber 
~\hspace{-7mm} \\
\label{eq:not}
\end{eqnarray}
where $p_i$ and $p_f$ are the (in general off-shell) momenta of the initial and final pions, respectively. 
The skewness can be written explicitly as
\begin{eqnarray}
\xi=\frac{p_i^+-p_f^+}{p_i^++p_f^+},
\end{eqnarray}
and satisfies $-1 \le \xi \le 1$.

In the partonic framework, $(x \pm \xi)P^+$ is the longitudinal momentum
carried by the initial (final) struck parton. The light-cone indices are defined in the convention $a^\pm= (a^0 \pm a^3)/\sqrt{2}$. 
To write the expression covariantly, one can introduce the null vector $n$ with the properties
\begin{equation}
P \cdot n = 1, \; q\cdot n = -2\xi, \; n^2=0.  \label{eq:n}
\end{equation}

The leading-twist chiraly even off-shell GPDs of the pion are defined in full analogy to the on-shell case~ (cf.~\cite{Diehl:2003ny}), namely
\begin{eqnarray}
&&\hspace*{1mm} \delta_{ab}\delta_{\alpha\beta}{H}^{0}(x,\xi,t,p_i^2,p_f^2)+i\epsilon^{abc}\tau^c_{\alpha\beta}{H}^{1}(x,\xi,t,p_i^2,p_f^2) = \nonumber \\
&&\hspace*{1mm} \!\!\frac{1}{2}\left.\int \! \frac{d z^-}{2\pi} e^{i x \, P^+ z^-} \langle\pi^b(p_f)|\overline{\psi}_\alpha(-\tfrac{z}{2}) \gamma^+ \psi_\beta(\tfrac{z}{2})|\pi^a(p_i)\rangle \right |_{\substack{z^+=0\\z^\perp=0}}, \nonumber \\ 
&& \hspace*{1mm} \delta_{ab}{H}^{g}(x,\xi,t,p_i^2,p_f^2)= \label{eq:Hdef}  \\
&&\hspace*{1mm} \!\! \left . \int \! \frac{d z^-}{2\pi P^+} e^{i x \, P^+ z^-} \langle\pi^b(p_f)| F^{+\mu}(-\tfrac{z}{2}) {F_\mu}^+(\tfrac{z}{2})  |\pi^a(p_i)\rangle \right |_{\substack{z^+=0\\z^\perp=0}}, \nonumber
\end{eqnarray}
where $\psi$ and $F^{\mu\nu}$ represent the quark and gluon fields. The squares of the incoming and outgoing pion momenta, $p_i^2$ and $p_f^2$, are not necessarily equal to $m_\pi^2$, and are treated as parameters. 
In the assumed light-cone gauge $A^+_a=0$, the Wilson link operators do not explicitly appear in the above definitions.
The subscripts $0,1$ denote the isospin of the quark GPDs. 
Indices $\alpha$ and $\beta$ represent the quark flavor, $a$, $b$ stand for the isospin of the pions, while $c$ is the isospin of the probing operator. 
The quark GPDs $H^{0,1}$ are related to the distributions of quarks and antiquarks as follows:
\begin{eqnarray}
&&{H}^{q,\bar{q}}(x,\xi,t,p_i^2,p_f^2) =\label{eq:Hq}  \\
&& ~~~~\frac{1}{2} \left [ {H}^{0}(x,\xi,t,p_i^2,p_f^2)\pm{H}^{1}(x,\xi,t,p_i^2,p_f^2) \right ], \nonumber
\end{eqnarray} 
By general arguments of the Lorentz covariance, the function ${H}^{q}(x,\xi,t,p_i^2,p_f^2)$ has the support $x \in [-|\xi |,1]$, whereas ${H}^{\bar{q}}(x,\xi,t,p_i^2,p_f^2)$ has the support $x \in [-1,|\xi |]$.\par

In general, the isovector ($I=1$) GPD is symmetric in $x$ while the isosinglet ($I=0$) and the gluon GPDs are anti-symmetric in $x$:
\begin{align}
    H^1(x,\xi,t,p_i^2,p_f^2) &= H^1(-x,\xi,t,p_i^2,p_f^2)\\
    H^0(x,\xi,t,p_i^2,p_f^2) &= -H^0(x,\xi,t,p_i^2,p_f^2).
\end{align}
This feature of the GPDs does not change whether the pion is on-shell or not. On the other hand, 
when the pion is off-shell, the GPDs loose the symmetry under time-reversal (or crossing), 
which is equivalent to the simultaneous replacement of $x \to -x$, $\xi \to -\xi$, or instead to the replacement  $p_f^2 \leftrightarrow p_i^2$. 
As a result,  the generalized form factors  $A^{s}_{j,i}$, defined as the $n^\text{th}$ moments  in the $x$ variable of the GPDs $H^s$ ($s=0,1,g$),
\begin{eqnarray}
&& \hspace{-7mm} \int_{-1}^1 dx \, x^j H^s(x,\xi,t,p_i^2,p_f^2)=\sum_{i=0}^{j+1} A^{s}_{j,i}(t,p_i^2,p_f^2) \xi^i   \label{eq:poly} \\
&& = A^{s}_{j,0}(t,p_i^2,p_f^2) +A^{s}_{j,1}(t,p_i^2,p_f^2) \xi+ \dots, \nonumber
\end{eqnarray}
are no longer even powers of $\xi$. Thus, though the polynomiality feature is retained, the odd powers of the skewness also enter the moments of the GPDs,
bringing a new set of form factors. Under the replacement $p_i^2 \leftrightarrow p_f^2$, we find that 
\begin{eqnarray}
A^{s}_{j,i}(t,p_f^2,p_i^2)=(-1)^i A^{s}_{j,i}(t,p_i^2,p_f^2).
\end{eqnarray}

\section{Electromagnetic and gravitational form factors\label{sec:ff}}

The first two lowest rank ($j=0$ and $1$ in Eq.~(\ref{eq:poly})) generalized (off-shell) form factors correspond
to the matrix elements of the conserved electromagnetic and energy-stress tensor currents. As such, they do not depend on the factorization
scale, and therefore are independent of the QCD evolution, which makes them particularly important objects.
Explicitly, one introduces
\begin{eqnarray}
&& N^1(\xi,t,p_i^2,p_f^2) \equiv \int_{-1}^1 dx \,  {H}^{1}= \label{eq:normvec}  \\ 
&& \hspace{2cm}2(F(t,p_i^2,p_f^2) - \xi G(t,p_i^2,p_f^2)),  \nonumber \\
&& N^0(\xi,t,p_i^2,p_f^2) \equiv \int_{-1}^1 dx \,  x[H^{0}+H^{g}]=\label{eq:normgrav} \\ 
&& \hspace{1cm}\theta_2(t,p_i^2,p_f^2) - 2\xi \theta_3(t,p_i^2,p_f^2) - \xi^2 \theta_1(t,p_i^2,p_f^2).  \nonumber
\end{eqnarray}
According to the above-mentioned symmetry arguments, the form factors multiplying 
even powers of $\xi$, namely $F$, $\theta_2$, and $\theta_1$, are even under the 
replacement $p_i^2 \leftrightarrow p_f^2$, whereas $G$ and $\theta_3$, corresponding to odd 
powers of $\xi$, are odd under $p_i^2 \leftrightarrow p_f^2$.

Our next task is to obtain general relations between various form factors of the pion using WTIs. 
Some important properties of the off-shell electromagnetic form factors were explored long ago in~\cite{Naus:1989em,Rudy:1994qb} and below we recall that methodology. 
In Appendix~\ref{app:WTIder} 
we review the derivations of the WTIs for the electromagnetic and gravitational form factors. 
It is customary to pass from the full (reducible or unamputated) vertices to the amputated vertices $\Gamma$, which appear as the building blocs of 
hadronic Feynman diagrams, together with the pion propagator $\Delta$. The leg amputation procedure for the electromagnetic and gravitational cases yields
\begin{eqnarray}
&&\Gamma^{\mu}(p_i,p_f)=\Delta^{-1}(p_i) G^{\mu}(p_i,p_f) \Delta^{-1}(p_f), \\
&&\Gamma^{\mu\nu}(p_i,p_f)=\Delta^{-1}(p_i) G^{\mu\nu}(p_i,p_f) \Delta^{-1}(p_f). \label{eq:ampute}
\end{eqnarray}

The WTI for the electromagnetic vertex of Eq.~(\ref{eq:preWTIEM}) assumes the form 
\begin{align}
q_\mu\Gamma^\mu(p_i,p_f) &= \Delta^{-1}(p_f^2) - \Delta^{-1}(p_i^2). \label{eq:wti1}
\end{align}
From the Lorentz covariance, the general form of the (positively charged) pion photon vertex is given by
\begin{align}
    \Gamma^\mu &= 2 P^\mu F(t,p_i^2,p_f^2) + q^\mu G(t,p_i^2,p_f^2), \label{eq:genEV}
\end{align}
hence 
\begin{align}
    q_\mu \Gamma^\mu &= (p_f^2-p_i^2) F(t,p_i^2,p_f^2) + t G(t,p_i^2,p_f^2), \label{eq:genEV1}
\end{align}
We note that when Eq.~(\ref{eq:genEV}) is contracted with $n_\mu$, it yields Eq.~(\ref{eq:normvec}).
By comparing Eq.~(\ref{eq:genEV1}) and~(\ref{eq:wti1}) one promptly finds~\cite{Naus:1989em,Rudy:1994qb} the relations
for the off-shell electromagnetic form factors
\begin{eqnarray}
 &&  \hspace{-6mm} G(t,p_i^2,p_f^2) = \frac{p_f^2-p_i^2}{t} \left( F(0,p_i^2,p_f^2) - F(t,p_i^2,p_f^2) \right)\!,~\label{eq:piG}\\
 &&  \hspace{-6mm} F(0,m_\pi^2,p^2) = F(0,p^2,m_\pi^2) = \frac{\Delta^{-1}(p^2)}{(p^2 - m_\pi^2)}, \label{eq:piF}\\
 &&  \hspace{-6mm} F(0,m_\pi^2,m_\pi^2) = 1. \label{eq:one}
\end{eqnarray}
The last equation is the charge sum rule (for the considered positively charged pion). It follows from Eq.~(\ref{eq:piF}) in the limit $p^2 \to m_\pi^2$ and the fact 
that for the canonical pion field in the vicinity of the pole $~\Delta(p^2)=1/(p^2-m_\pi^2)+{\cal O}(1)$.
Moreover,
\begin{eqnarray}
G(0,p_i^2,p_f^2) =(p_i^2-p_f^2)dF(t,p_i^2,p_f^2)/dt |_{t=0}. \label{eq:piG2}
\end{eqnarray}

Similar arguments carry over to the case of the stress-energy tensor, as argued in~\cite{Broniowski:2022iip}. 
The WTI for the amputated gravitational vertex(Eq.~(\ref{eq:ampute})) follows from Eq.~(\ref{eq:preWTIGFF}) and takes the form
\begin{eqnarray}
&& \hspace*{-4mm} q_\mu \Gamma^{\mu\nu} (p_f,p_i) = p_i^{\nu} \Delta^{-1}(p_f^2) - p_f^\nu\Delta^{-1}(p_i^2)  =  \label{eq:wti2} \\
&&\hspace*{-4mm} ~~~P^{\nu} [\Delta^{-1}(p_f^2) - \Delta^{-1}(p_i^2)]-\tfrac{1}{2} q^\nu [\Delta^{-1}(p_f^2) + \Delta^{-1}(p_i^2)]. \nonumber
\end{eqnarray}
The most general vertex allowed by the symmetries is
\begin{eqnarray}
 \Gamma^{\mu\nu} &=& \tfrac{1}{2}\left[ (q^2 g^{\mu\nu} -q^\mu q^\nu)\theta_1 + 4 P^\mu P^\nu \theta_2  \right .+ \nonumber \\
&& ~~~\left . (q^\mu P^\nu + P^\mu q^\nu) \theta_3 - g^{\mu\nu}\theta_4 \right] \label{eq:grff}
\end{eqnarray}
where, $\theta_i\equiv \theta_i(t,p_i^2,p_f^2)$ are the gravitational form factors.
This form generalizes the on-shell definition of~\cite{Donoghue:1991qv}. Upon the contraction with $q_\mu$ we get
\begin{eqnarray}
&& \hspace*{0mm} q_\mu \Gamma^{\mu \nu} = (p_f^2\!-\!p_i^2) P^\nu \theta_2 +[tP^\nu +  \tfrac{1}{2}(p_f^2\!-\!p_i^2) q^\nu] \theta_3 -  \tfrac{1}{2}q^\nu \theta_4. \nonumber \\ 
&&  \label{eq:contr}
\end{eqnarray}
Since the four-vectors $P$ and $q$ are linearly independent, comparing Eqs.~(\ref{eq:wti2}) and (\ref{eq:contr}) we find two independent relations
\begin{eqnarray}
&& \hspace{-7mm} (p_f^2\!-\!p_i^2) \theta_2 + t  \theta_3 =  \Delta\!^{-1}(p_f^2) - \Delta\!^{-1}(p_i^2), \label{eq:rel1} \\
&& \hspace{-7mm} (p_f^2\!-\!p_i^2) \theta_3-\theta_4 =  -[\Delta\!^{-1}(p_f^2) + \Delta\!^{-1}(p_i^2)]. \label{eq:rel2}
\end{eqnarray}

The form factor $\theta_1$, multiplying a transverse tensor, remains unconstrained by the WTI relations. 
Contraction of Eq.~(\ref{eq:grff}) with $n_\mu n_\nu$ yields the right-hand side of Eq.~(\ref{eq:normgrav}) multiplied by a factor of $\tfrac{1}{2}$, which originates form the definition~(\ref{eq:Hdef}).
Note that $\theta_4$ does 
not enter Eq.~(\ref{eq:normgrav}) because $n^2=0$, hence it is not accessible in studies of the GPDs.

From Eq.~(\ref{eq:rel1})  
\begin{eqnarray}
 &&  \hspace{-7mm} \theta_3(t,p_i^2,p_f^2) = \frac{p_f^2\!-\!p_i^2}{t} \left [  \theta_2(0,p_i^2,p_f^2) -  \theta_2(t,p_i^2,p_f^2)\right ], \label{eq:theta3}
\end{eqnarray}
with
\begin{eqnarray}
\theta_2(0,m_\pi^2,p^2) = \theta_2(0,p^2,m_\pi^2) = \frac{\Delta\!^{-1}(p^2)}{(p^2-m_\pi^2)}, \label{eq:theta20}
\end{eqnarray}
and the mass sum rule
\begin{eqnarray}
\theta_2(0,m_\pi^2,m_\pi^2)=1.
\end{eqnarray} 
From Eq.~(\ref{eq:theta3}) we immediately find that
\begin{eqnarray}
\theta_3(0,p_i^2,p_f^2) =(p_i^2\!-\!p_f^2)d\theta_2(t,p_i^2,p_f^2)/dt |_{t=0}.
\end{eqnarray}
The above relations for the off-shell gravitational form factors mirror one-to-one those for the off-shell electromagnetic form factors. 

Interestingly, we also get a relation between the off-shell electromagnetic and gravitational form factors at $t=0$, namely
\begin{eqnarray}
\hspace{-7mm}\theta_2(0,p_i^2,p_f^2)=F(0,p_i^2,p_f^2) =  \Delta^{-1}(p_f^2) - \Delta^{-1}(p_i^2).   \label{eq:F2}
\end{eqnarray} 
This important relation will be used in Sec.~\ref{sec:offpi} to argue that the off-shell effects largely cancel at low $\xi$ in the Compton form factor entering the pion electroproduction processes. 

Finally, using Eq.~(\ref{eq:rel2}) we can express $ \theta_4$ solely in terms of  $\theta_2$, namely
\begin{eqnarray}
 && \hspace*{1mm}   \theta_4(t,p_i^2,p_f^2) = \frac{(p_f^2\!-\!p_i^2)^2}{t} \left [  \theta_2(0,p_i^2,p_f^2) -  \theta_2(t,p_i^2,p_f^2)\right ] \nonumber \\
 &&\hspace*{1mm}  ~~~~ + (p_i^2-m_\pi^2) \theta_2(0,p_i^2,m_\pi^2)+ (p_f^2-m_\pi^2) \theta_2(0,m_\pi^2,p_f^2). \nonumber \\ \label{eq:theta4}
\end{eqnarray}

All the above relations hold in general in theories where the pion satisfies the PCAC relation, needed to derive the WTIs. 
More discussion on the assumptions and the derivations of WTIs are presented in Appendix~\ref{app:WTIder}.

\section{Off-shell GPDs of the pion in the Spectral Quark Model \label{sec:model}}

The results presented up to now for the GPDs and the related form factors were completely general and followed from the definitions and symmetries.
In the subsequent parts of this paper we extensively illustrate these features within a chiral quark model, where the pion is a composite field (satisfying PCAC).
The purpose of an explicit realization is twofold: first, intricate features are displayed within a model, which albeit simple, leads to expressions for the GPD which are 
quite involved and nontrivial. In particular, they exhibit no factorization in $x$, $t$, $\xi$, and the off-shell momenta. The other reason is to obtain a model estimate for the 
size of the off-shellness effects as expected in electroproduction process. Since the applied model is realistic in other predictions for the pion, one can expect it may provide a useful guideline also for the physical phenomena studied here, such as the Sullivan process. 

\subsection{Model}

The model we use has the Lagrangian density with a nonlinear pion field realization,
\begin{eqnarray}
{\cal L}(x)=\bar{\psi}(x) \left [ i \slashed{\partial} - \omega e^{i \gamma_5  \tau^a \phi^a(x)/f} \right ] \psi(x), \label{eq;lag}
\end{eqnarray}
where $\psi$ is the quark field, $\omega$ is the quark mass, $\tau_a$ are the Pauli isospin matrices, $\phi^a$ is the pion field, and $f$ is the pion decay constant. 
For the physical pion mass $f=93$~MeV, while in the chiral limit $f=86$~MeV.  
With the nonlinear realization one avoids introducing the $\sigma$ field of the linear model, which leads to somewhat more complicated (but qualitatively equivalent) results. 
The interactions in the model are local (point-like), and the pion is a pseudo-Goldstone boson according the the Nambu--Jona-Lasinio mechanism.

\subsection{GPDs at the one-quark-loop level}

\begin{figure}[tb]
\centering
\includegraphics[width=0.35\textwidth]{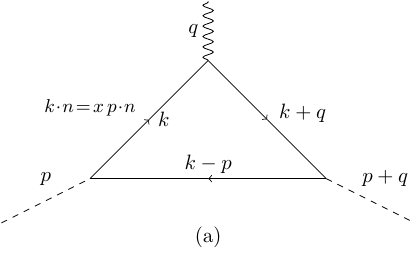} \\
\vspace{3mm}
\includegraphics[width=0.35\textwidth]{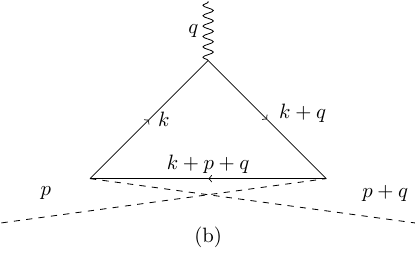}\\
\vspace{3mm}
\includegraphics[width=0.15\textwidth]{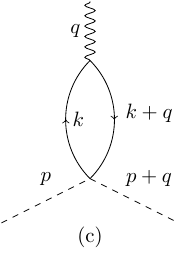}
\caption{One-loop Feynman diagrams for the model evaluation of the GPDs. \label{fig:feyndia}}
\end{figure}

The one-quark-loop Feynman diagrams appearing in our calculations of the GPDs are shown in Fig.~\ref{fig:feyndia}. They follow directly from the definitions (Eq.~(\ref{eq:Hdef})).
The fermion lines correspond to constituent quarks, whose large mass, denoted at $\omega$, follows from the spontaneous chiral symmetry breaking~\cite{RuizArriola:2002wr}. 
Diagram~(c) results from from expanding the interaction term in Eq.~(\ref{eq;lag}) to second order in $\phi_a$.
With the standard Feynman rules in the momentum representation, the diagrams of  Fig.~\ref{fig:feyndia} yield the following contributions to the off-shell pion GPDs:
\begin{eqnarray}
&&\hspace{-4mm} H_a(x,\xi,t,p_i^2,p_f^2)=  \frac{N_c w^2}{f^2}\int \frac{d^4 k}{(2\pi)^4} \delta(k\cdot n-x)\times \label{eq:ol}\\
&&~~~~{\rm Tr}\left [ \gamma \cdot n S(k-\tfrac{1}{2}q) \gamma_5 S(k-P) \gamma_5  S(k-\tfrac{1}{2}q) \right ], \nonumber \\
&&\hspace{-4mm} H_b(x,\xi,t,p_i^2,p_f^2)= \frac{N_c w^2}{f^2}\int \frac{d^4 k}{(2\pi)^4} \delta(k\cdot n-x)\times \\
&&~~~~{\rm Tr}\left [ \gamma \cdot n S(k-\tfrac{1}{2}q) \gamma_5 S(k+P) \gamma_5  S(k-\tfrac{1}{2}q) \right ], \nonumber \\
&&\hspace{-4mm} H_c(x,\xi,t,p_i^2,p_f^2)=  \frac{N_c w}{f^2}\int \frac{d^4 k}{(2\pi)^4} \delta(k\cdot n-x)\times \\
&&~~~~{\rm Tr}\left [ \gamma \cdot n S(k-\tfrac{1}{2}q) i  S(k-\tfrac{1}{2}q) \right ], \nonumber
\end{eqnarray}
where the quark propagator is 
\begin{eqnarray}
S(l)=\frac{i}{\slashed{l}-\omega +i \epsilon}.
\end{eqnarray}
In the assumed chiral limit, the quark-meson coupling constant is equal to $\omega/f$.
The isospin decomposition yields
\begin{eqnarray}
&&\hspace*{-4mm} H^1(x,\xi,t,p_i^2,p_f^2)=H_a(x,\xi,t,p_i^2,p_f^2)-H_b(x,\xi,t,p_i^2,p_f^2), \nonumber \\
&&\hspace*{-4mm} H^0(x,\xi,t,p_i^2,p_f^2)=\label{eq:is} \\
&&\hspace*{-4mm} H_a(x,\xi,t,p_i^2,p_f^2)+H_b(x,\xi,t,p_i^2,p_f^2)+H_c(x,\xi,t,p_i^2,p_f^2). \nonumber
\end{eqnarray}

\subsection{Spectral regularization}

The Nambu--Jona-Lasinio model and its descendants are low-energy models and thus, by physics arguments, a cut-off (regularization) 
of the hard momenta needs to be introduced.
In fact, the diagrams of Fig.~(\ref{fig:feyndia}) are logarithmically divergent. When introducing a regularization, care is needed not to spoil the the symmetries of the 
theory, such as the Lorentz and gauge invariances, or preservation of anomalies~\cite{RuizArriola:2002wr}. There are several ways of doing this in a proper way. Popular schemes are the 
Pauli-Villars or proper-time regularizations. 

In this paper we apply the spectral regularization~ \cite{RuizArriola:2003bs}, which amounts to overlaying 
contributions from different quark masses in the quark loop, similar in spirit to the 
Pauli-Villars method. In the Spectral Quark Model (SQM), however, the quark masses are distributed in the complex plane 
and the evaluation of the amplitudes involves the following operation \cite{RuizArriola:2003bs,Broniowski:2007si},
\begin{align}
    \mathcal{A}^\text{SQM} &= \int_C d\omega  \rho(\omega) \mathcal{A}^\text{bare}
\end{align}
where, $\mathcal{A}^\text{bare}$ represents the unregularized amplitude, $\rho(\omega)$ is the regularizing spectral function, 
and $C$ is a suitable contour of integration (cf. Fig.~1 in \cite{RuizArriola:2003bs}.

One of the main advantages of SQM is a possibility of an exact implementation of the vector meson dominance (VMD) in the (on-shell) pion electromagnetic form factor, 
\begin{align}
    F_V(t) &= \frac{M_V^2}{M_V^2-t}, \label{eq:vmds}
\end{align}
where $M_V\sim m_\rho \sim 770$~MeV is the vector meson mass. In the assumed chiral limit, it is the only parameter of the model.
The form(Eq~\ref{eq:vmds}) is accomplished by choosing the spectral function in the form~\cite{RuizArriola:2003bs},
\begin{align}
    \rho(\omega) &= \frac{1}{2\pi i}\frac{1}{\omega}\frac{1}{\left(1-4\omega^2/M_V^2 \right)^{5/2}}, 
\end{align}
with the  consistency relation $M_V^2 = 24 \pi^2 f^2 /N_c$ which works
well phenomenologically.  SQM has been successfully applied to evaluate
such properties of the pion as its electromagnetic, gravitational and
higher order form factor, structure function, PDF, GPDs, transition
form factor, decay constant, and so
on~\cite{RuizArriola:2003bs,Broniowski:2003rp,Dorokhov:2006qm,Broniowski:2007fs,Broniowski:2007si,Broniowski:2008hx}.

\subsection{GPDs at the quark-model scale}

The GPDs are evaluated using the above-described model at the one-quark-loop level. 
The calculation provides the leading-$N_c$ quark model result which holds at the quark model scale, 
which then must be evolved to higher experimental of lattice scales~~\cite{Davidson:1994uv,Broniowski:2007si}, as described in Sec.~\ref{sec:evol}. 
The formalism and technicalities are detailed in Refs.~\cite{Broniowski:2007si} and~\cite{Shastry:2022obb}. 

One can write down a decomposition of the model one-quark-loop amplitudes in the form Eq.~(\ref{eq:ol})
\begin{eqnarray}
&&\hspace*{1mm} H^{1,0}(x,\xi,t,p^2)= \label{eq:hgen} \\
&&\hspace*{1mm}  ~~\tfrac{1}{2} \left [ (1-\xi) I(x,\xi,p_i^2) + (1+\xi) I(x,-\xi,p_f^2) + \right . \nonumber \\
&&\hspace*{1mm}  ~~~ + \left . [m_f^2 (\xi +x)+m_i^2 (x-\xi )+t(1- x)] J(x,\xi,t,p_i^2,p_f^2) \right] \nonumber \\ && \hspace{6cm}\pm (x \leftrightarrow -x), \nonumber 
\end{eqnarray}
with the one-loop functions $I$ (two-point) and $J$ (three-point) provided in Appendix~\ref{app:IJ}. Note that the $x$, $\xi$,  $t$, $p_i^2$ and $p_f^2$ 
dependence in general does not factorize.
While the explicit form of the general formulas (Eq.~(\ref{eq:hgen})) is long, they become very simple for the case where $p_i^2=p_f^2=t=0$, with
the step-wise functions~\cite{Broniowski:2007si},
\begin{eqnarray}
&&    H^{1}(x,\xi,0,0,0) =  \Theta[(x+1)(1-x)], \\
&&    H^0(x,\xi,0,0,0)  = \Theta[(x-\xi)(1-x)] - (x \leftrightarrow -x). \nonumber 
\end{eqnarray}
Another very simple case occurs for the  half-off-shell PDF case 
\begin{eqnarray}
p_i^2=p^2, \;\;\;p_f^2=0,
\end{eqnarray}
with $t=0$ and $\xi=0$:
\begin{align}
    H^{1,0}(x,0,0,p^2,0) &= \frac{\Theta(x(1-x))}{\left (1 - \frac{4 x (1 - x)p^2}{M_V^2} \right)^{3/2}} \pm  (x \leftrightarrow -x). \label{eq:H000}
\end{align}

\begin{figure}[t]
    \centering
    \includegraphics[width=0.38\textwidth]{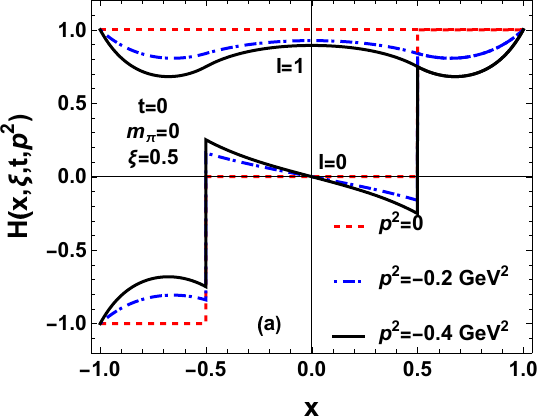}\\\vspace{3mm}\includegraphics[width=0.38\textwidth]{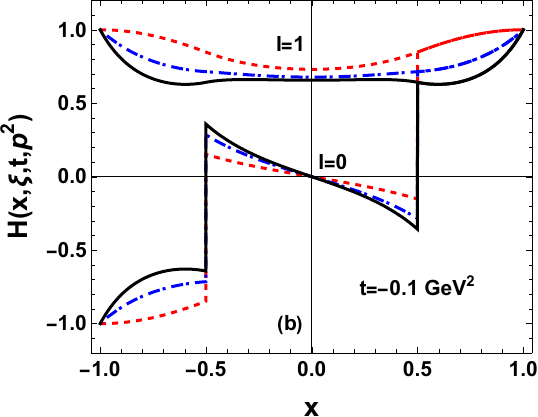}
    \caption{Half-off-shell pion GPDs at $\xi=0.5$ for (a)~$t=0$ and (b)~$t=-0.1$~GeV$^2$, evaluated in the chiral limit in SQM at the quark model scale for several values 
    of the off-shell parameter~$p^2$.}
    \label{fig:GPDin}
\end{figure}

Although in SQM one can compute analytically the fully off-shell
formulas for the pion GPDs, they are lengthy and not illuminating. We
thus focus on the half-off-shell results (relevant {\em i.a.} for the
Sullivan process) in the chiral limit, where the expressions are
simpler.  The GPDs obtained using
SQM for $t=0$ and $t=-0.1$ GeV$^2$, and for various representative
values of the off-shellness $p^2$ are plotted in
Fig.~\ref{fig:GPDin}~(a similar range of numerical values has been considered in~\cite{Qin:2017lcd}). One immediately notices that the effect of
the off-shellness is quite pronounced even for moderate values of
$p^2=-0.2~\text{ GeV}^2$. Furthermore, the DGLAP region $|x|>|\xi|$ is
affected more than the ERBL region ($|x|<|\xi|$).

In SQM, the normalization of the GPDs following from Eqs.~(\ref{eq:normvec},\ref{eq:normgrav}) (see also Sec.~\ref{sec:vgff}) is
\begin{eqnarray}
&& \hspace{-5mm}  \int_{-1}^1 dx \,  {H}^{1} \equiv N_1(\xi,t,p^2)=2 \frac{M_V^2}{M_V^2-t}\frac{M_V^2-\xi  p^2}{p^2-M_V^2 } = \label{eq:normvec2}\\
&& \hspace{1.7cm}\frac{M_V^2}{M_V^2-t} \left [  1+\frac{(1-\xi ) p^2}{M_V^2}+{\cal O}\left(\frac{p^4}{M_V^4}\right) \right ], \nonumber
\end{eqnarray}

\begin{eqnarray}
&&\hspace*{1mm}  \int_{-1}^1 dx \,  x[H^{0}+H^{g}] \equiv N_0(\xi,t,p^2) = \label{eq:normgrav2} \\
&& \frac{M^2 (\xi -1) \left(\frac{(\xi -1) p^2 \left(p^2-t\right)}{M^2-p^2}-L \left(-2
   \xi  p^2+\xi  t+t\right)\right)}{\left(p^2-t\right)^2} =\nonumber \\
&&\hspace*{1mm} 1-\xi ^2 +\frac{(1-\xi ^2) t}{2M_V^2}- \frac{(1-\xi)\xi p^2 }{M_V^2}+  {\cal O}\left(\frac{t^2}{M_V^4},\frac{p^2}{M_V^4},\frac{p t}{M_V^4} \right). \nonumber
\end{eqnarray}
This implies, as expected, that the normalizations $N_{0,1}$ acquire corrections from the off-shellness, which at small $p^2/M_V^2$ behave as $(1-\xi)p^2/M_V^2$. The relevant scale here is the vector meson mass (which is the only scale in our model), whereas the magnitude is proportional to $(1-\xi)$. 

In Fig. \ref{fig:GPDin} we plot the pion GPDs at the quark model scale, whose behavior reflects the above normalization conditions. 
We observe that in the DGLAP regions in~Fig. \ref{fig:GPDin}, the value of the GPDs can reduce by almost $30-40\%$ at $|p^2| \sim M_V^2$ compared to the on-shell value, with a strongest effect for $|x|$ near $\xi$.

\subsection{QCD evolution \label{sec:evol}}

\begin{figure}[t]
    \centering
    \includegraphics[width=0.38\textwidth]{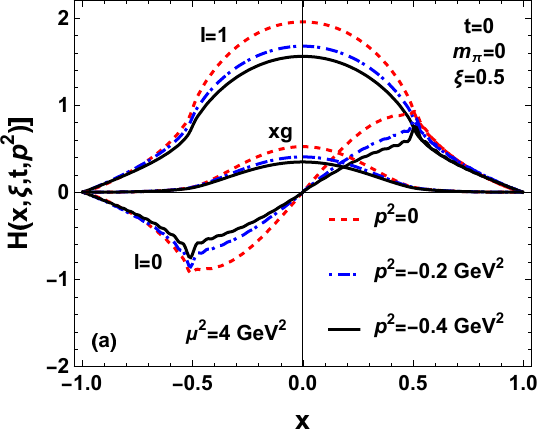}\\ \includegraphics[width=0.38\textwidth]{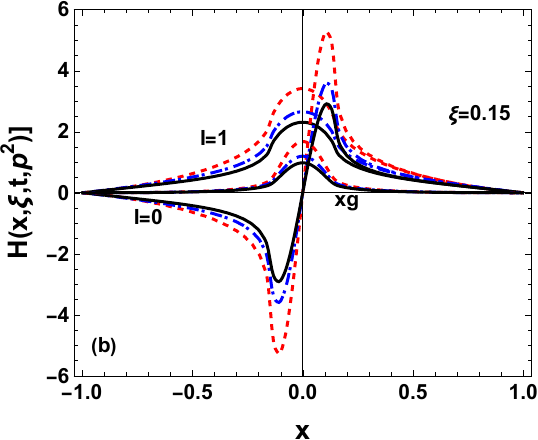}
    \caption{Half-offshell pion GPD for $t=0$ at $\xi=0.5$ and $\xi=0.15$, evolved to $Q^2=4~\text{GeV}^2$ with LO DGLAP-ERBL equations.}
    \label{fig:GPDev}
\end{figure}

The GPDs are scale dependent objects which undergo the DGLAP-ERBL QCD
evolution equations~\cite{Muller:1994ses,Diehl:2003ny}.  We note that
off-shellness of the initial and final hadronic states (pions) does
not affect the QCD evolution kernel in the assumed Bjorken limit, so
the method proceeds in the usual way.  The GPDs obtained from SQM
  hold at the quark model scale of $\mu_q^2 \sim 0.1~\text{ GeV}^2$ and
serve as the initial conditions for the evolution.\footnote{This
    initial quark model scale produces very similar results for LO and
    NLO evolution of the on-shell PDFs, as shown in~\cite{Davidson:2001cc}.}  We evolve the GPDs
to a representative scale of $\mu^2=4$~GeV$^2$ using the procedure
detailed in~\cite{Golec-Biernat:1998zbo}. The evolved GPDs are
displayed in Fig.~\ref{fig:GPDev}.  The above described characteristic
effects resulting from the off-shellness of the incoming pion remain.
Further, similar features are exhibited by the gluon distributions as
well. We note the known phenomenon that the evolution smooths out the
quark-model initial condition; it causes the GPDs to vanish at the
end-points $x=0,1$, as well as makes them continuous at $x=\pm \xi$.

For $\mu^2\to \infty$, the GPDs tend to the asymptotic forms
with the support in the ERBL region only. Explicitly,
\begin{eqnarray}
\label{eq:asymptotics}
&&H^{1}(x,\xi,t,p_i^2,p_f^2)= N^1(t,\xi,p^2,p_f^2) \frac{3}{4|\xi |} \left(1\!-\!\frac{x^2}{\xi^2}\right), \nonumber \\
&&H^{0}(x,\xi,t,p_i^2,p_f^2)= \nonumber \\ && ~~~~ N^0(t,\xi,p^2,p_f^2) \frac{15}{4 \xi^2} \frac{N_f}{4 C_F\!+\!N_f} \frac{x}{\xi}\left(1\!-\!\frac{x^2}{\xi^2}\right), \nonumber \\
&&x H^g(x,\xi,t,p_i^2,p_f^2)=  \nonumber \\ && ~~~~N_0(t,\xi,p^2,p_f^2) \frac{15}{16 |\xi|} \frac{4C_F}{4C_F\!+\!N_f} \left(1\!-\!\frac{x^2}{\xi^2}\right)^2, \label{eq:asev}
\end{eqnarray} 
where $N_f$ is the number of flavors, $C_F=\tfrac{4}{3}$, and the moments (Eqs.~(\ref{eq:normvec},\ref{eq:normgrav})) appear in the normalization.

\begin{figure}[t]
    \centering
    \includegraphics[width=0.38\textwidth]{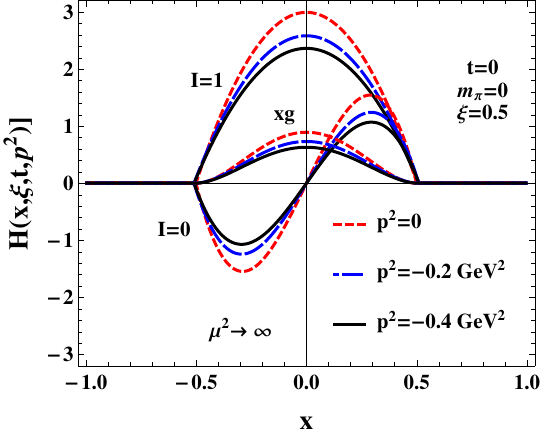}\
%    \\includegraphics[scale=0.58]{eapt1.pdf}
    \caption{Asymptotic half-offshell pion GPDs at $t=0$ and $\xi=0.5$ for various values of the off-shell parameter $p^2$.}
    \label{fig:GPDasym}
\end{figure}

The qualitative features of the dependence on $p^2$ of the half-off-shell GPDs in the ERBL region are preserved in the asymptotic limit, as can be seen from  Fig.~\ref{fig:GPDasym}. The symmetric and antisymmetric GPDs of the pion remain equal to each other for $x>\xi$ when evolved to energies above the quark model scale, where as the differences continue to show in the ERBL region. In the asymptotic limit, the GPDs go to zero in the DGLAP region. In the ERBL region, the $I=1$ GPD is quadratic as seen from Eq.~(\ref{eq:asev}). The dependence of the asymptotic GPDs on the off-shellness resides in the normalization factors, which in turn are the moments of the GPDs. 

\section{Form factors in the Spectral Quark Model \label{sec:vgff}}

We now turn our attention to the lowest moments of the GPDs defined in Eq.~(\ref{eq:normvec}) and Eq.~(\ref{eq:normgrav}), 
evaluated in SQM in the chiral limit for the half-off-shell case. We shall also need the inverse 
pion propagator in SQM, which in the chiral limit is equal to 
\begin{eqnarray}
\Delta^{-1}(p^2)=\frac{p^2 M_V^2}{M_V^2-p^2} \equiv \frac{p^2}{Z(p^2)}, \label{eq:invpiprop}
\end{eqnarray}
where $Z(p^2)=(M_V^2-p^2)/M_V^2$ is the pion field renormalization, with $Z(0)=1$.

\subsection{Vector form factors \label{sec:vf1}}

\begin{figure}[t]
    \centering
    \includegraphics[width=0.4\textwidth]{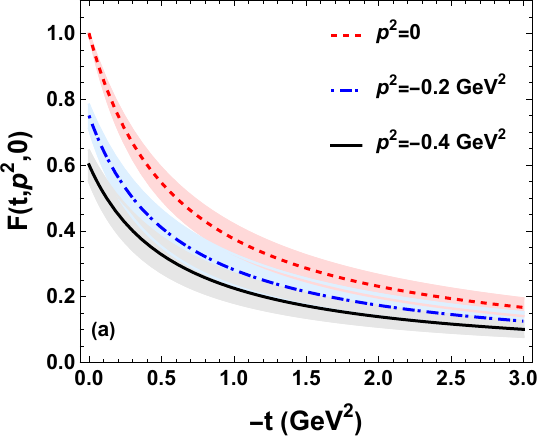}\\\includegraphics[width=0.4\textwidth]{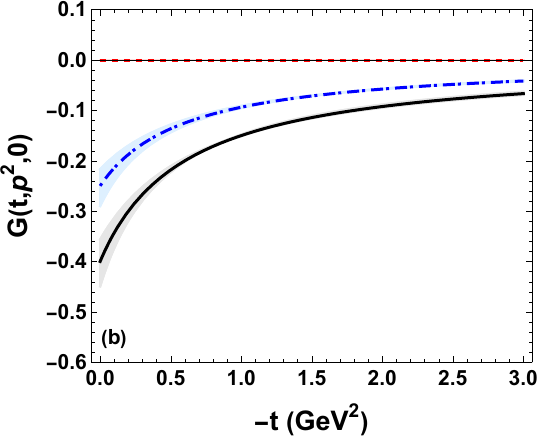}
    \caption{Half off-shell electromagnetic form factors of the pion in SQM.
    The lines correspond to $M_V=776$~MeV, whereas the bands indicate the uncertainty due to the width of the $\rho$ meson, $\Gamma_\rho=150$~MeV.
    \label{fig:FFvec}}
\end{figure}

The expressions for $F$ and $G$ have a particularly simple structure, exhibiting (in the assumed chiral limit) 
a factorized dependence on $t$ and $p^2$~\cite{Broniowski:2022iip},
\begin{align}
F(t,p^2,0) &= \frac{M_V^4}{\left(M_V^2-p^2\right) \left(M_V^2-t\right)}, \label{eq:ffvecF} \\
G(t,p^2,0) &=\frac{p^2 M_V^2}{\left(M_V^2-p^2\right) \left(M_V^2-t\right)}. \label{eq:ffvecG}
\end{align}
In the special case we have $F(0,p^2,0)=Z^{-1}(p^2)$, in agreement with the general relation Eq.~(\ref{eq:piF}).
They are plotted in Fig.~\ref{fig:FFvec} for three representative values of $p^2$. 
The central lines correspond to $M_V=775$~MeV, while the bands indicate the 
width of the $\rho$ meson resonance, $\Gamma_\rho=150$~MeV.

As is well known, in the on-shell case the vector meson dominance, built in by construction in SQM, reproduces well the experimental data at moderately low values of $t$~\cite{Amendolia:1986wj,Volmer:2000ek}. It also reproduces the results of lattice simulations~\cite{Brommel:2006ww,Hackett:2023nkr}.
Naturally, the form of Eqs.~(\ref{eq:ffvecF},\ref{eq:ffvecG}) complies to the general relations following from WTI, Eqs.~(\ref{eq:piG}-\ref{eq:piG2}).
We stress that at low $p^2$ the dependence on the off-shellness is $\sim p^2/M_V^2$, which should be considered a significant effect.

We draw attention to the off-shell electromagnetic form factors of the pion extracted in the framework of the chiral 
perturbation theory ($\chi$PT)~\cite{Rudy:1994qb,Koch:2001ii}. 
The one-loop $\chi$PT result shows a linear dependence of $G$ on $p^2$, with the leading coefficient equal to $2L_9/F^2$ (at $t=0$). 
This matches the leading term in the expansion of Eq.~(\ref{eq:ffvecG}) in $p^2$ when the SQM value for $L_9$ is 
used~\cite{Megias:2004uj} $L_9=\frac{N_c}{48\pi^2}$. Together with the relation~(\ref{eq:f}) this indeed yields $2L_9/f^2=1/M_V^2$.

\subsection{Gravitational form factors \label{sec:gf1}}

The explicit expressions for the four half-off-shell gravitational form factors  in SQM in the chiral limit are
\begin{align}
\theta_1(t,p^2,0) &=\frac{M_V^2 \left[\frac{p^2 (t-p^2)}{M_V^2-p^2}+(t-2 p^2) L \right]}{\left(t-p^2\right)^2}, \label{eq:ffgrav} \\
\theta_2(t,p^2,0) &= \frac{M_V^2 \left[\frac{p^2 (p^2-t)}{M_V^2-p^2}+t L\right]}{\left(t-p^2\right)^2}, \nonumber \\
\theta_3(t,p^2,0) &= \frac{p^2 M_V^2 \left[p^2-t+(M_V^2-p^2)L \right]}{\left(t-p^2\right)^2
   \left(M_V^2-p^2\right)}, \nonumber \\
\theta_4(t,p^2,0) &=   
   \frac{p^2 M_V^2 \left[\left(p^2\!-\!t\right) (2 p^2\!-\!t)+p^2 (M_V^2-p^2) L\right]}{(t-p^2)^2 (M_V^2-p^2)}, \nonumber
\end{align}
where  $L=\log\frac{M_V^2-p^2}{M_V^2-t}$. 
One can promptly verify that these expressions satisfy the general conditions given in Eqs.~(\ref{eq:theta3}-\ref{eq:theta4}) following from the gravitational WTI. 

Unlike the case of the half-off-shell electromagnetic form factors, form factors of Eq.~(\ref{eq:ffgrav}) do not exhibit factorization in $t$ and $p^2$. 
Their  expansion up to linear terms in $p^2$ and $t$ is 
\begin{align}
\theta_1(t,p^2,0) &=1+\frac{t}{2M_V^2} + \cdots, \label{eq:ffgrave} \\
\theta_2(t,p^2,0) &= 1+\frac{t}{2M_V^2}+\frac{p^2}{M_V^2} + \cdots , \nonumber \\
\theta_3(t,p^2,0) &= \frac{p^2}{2M_V^2}+ \cdots , \nonumber \\
\theta_4(t,p^2,0) &= p^2 + \cdots. \nonumber
\end{align}
In the opposite ($t\to\infty$) limit one has 
\begin{align}
\theta_1(t,p^2,0) &\sim\theta_2(t,p^2,0)\sim \label{eq:ffgraveas} \\
 &\frac{M_V^2}{t} \left(\log \left(\frac{p^2-{M_V}^2}{t}\right)+\frac{p^2}{{M_V}^2-p^2}\right), \nonumber \\
\theta_3(t,p^2,0) &= \frac{M_V^2 p^2}{t(p^2-M_V^2)}, \nonumber \\
\theta_4(t,p^2,0) &= \frac{M_V^2 p^2 (p^2-t)}{t(p^2-M_V^2)}. \nonumber
\end{align}

\begin{figure}[ht]
    \centering
    \includegraphics[width=0.38 \textwidth]{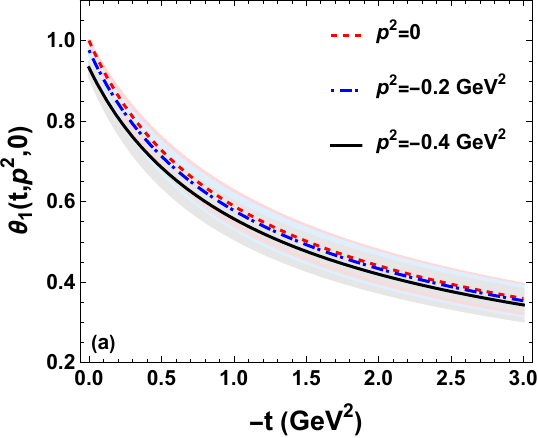} \\
    \includegraphics[width=0.38 \textwidth]{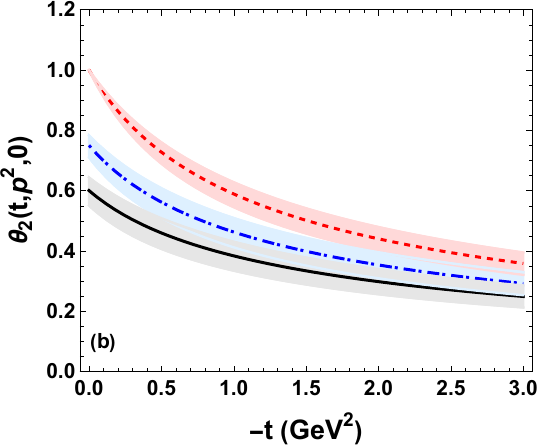} \\ 
    \includegraphics[width=0.38 \textwidth]{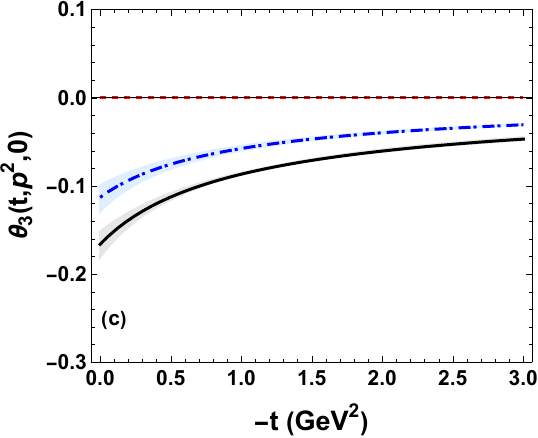}%
    \caption{Same as in Fig.~\ref{fig:FFvec}, but for the half-offshell gravitational form 
    factors $\theta_1$, $\theta_2$, and $\theta_3$. \label{fig:FFg12}}
\end{figure}

The value of the on-shell gravitational form factor $\theta_2$ at $t=0$ represents the mass sum rule for the pion and is equal to $1$~\cite{Donoghue:1991qv}. However, as 
follows from the general relation~\ref{eq:rel1}, for the off-shell case  $\theta_2(0,p^2,0)=Z^{-1}(p^2)$.
The derivative of the on-shell $\theta_2$ at the origin provides the mass radius $r_m$ of the pion. In the SQM, from Eq.~(\ref{eq:ffgrav}), we get,
\begin{align}
    \langle r_m^2 \rangle &= \frac{1}{6}\frac{d\theta_2(t,p^2,0)}{dt}\Big|_{t=0}=\frac{1}{12M_V^2}.
\end{align}
Note that the mean square mass radius of the pion is half of its electromagnetic counterpart in the on-shell limit \cite{Broniowski:2008hx}. However, this ratio reduces as one of the pions becomes off-shell. The ratio of the mean square charge radii is thus given by~\cite{Broniowski:2008hx}
\begin{align}
    \frac{\langle r_E^2 \rangle}{\langle r_m^2 \rangle} = 2.
\end{align}

The half-off-shell form factors $\theta_1$,  $\theta_2$, and $\theta_3$ from SQM are plotted in Fig.~\ref{fig:FFg12}.
We note that the effect of off-shellness on $\theta_1$ is small, $\sim6\%$ when $t=0$ 
and $\sim 3\%$ when $t=-0.1~\text{GeV}^2$ for $-p^2$ in the range $0 - 0.4~\text{ GeV}^2$.  In the case of $\theta_1$, 
the logarithmic term cancels most of the off-shell contributions coming from the rest of the expression, whereas this is not the case in  $\theta_2$. 
Thus, $\theta_2$ exhibits a stronger dependence on $p^2$. 
We note that $\theta_3$ decreases as  $1/t$ asymptotically, whereas $\theta_4$ tends to a constant.

\begin{figure}[t]
\centering 
 \includegraphics[width=0.38\textwidth]{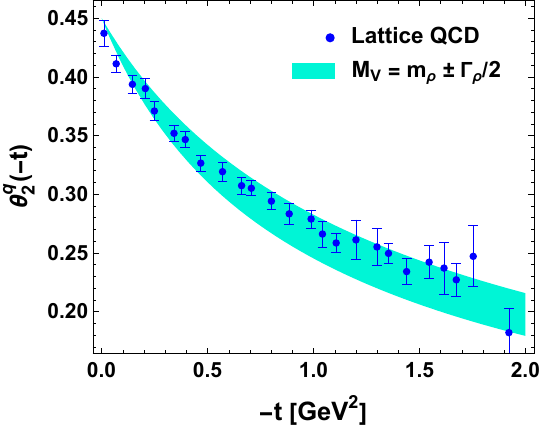}
 \caption{Comparison of the quark part of the on-shell gravitational factor factor $\theta_2$ with the lattice data of Ref.~\cite{Hackett:2023nkr}. The model value at the origin
 follows from the QCD evolution to $\mu^2=2~{\rm GeV}^2$~\cite{Broniowski:2008hx}. The 
 model band represents the width of the $\rho$ meson, $\Gamma_\rho=150~{\rm MeV}$. \label{fig:ffcomplatt}}
\end{figure}

In Fig.~\ref{fig:ffcomplatt} we show a comparison of $\theta_2$ obtained in SQM with a recent
lattice extraction~\cite{Hackett:2023nkr}, which are in good agreement. This further buttresses the assumption of the meson dominance implemented in SQM 
and its applicability to the gravitational form factors. 

\section{Compton form factors \label{sec:cff}}

\begin{figure}[t]
    \centering
    \includegraphics[width=0.38 \textwidth]{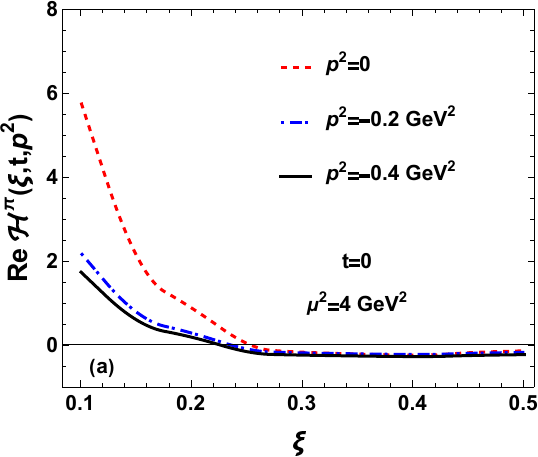}\\ 
    \includegraphics[width=0.38 \textwidth]{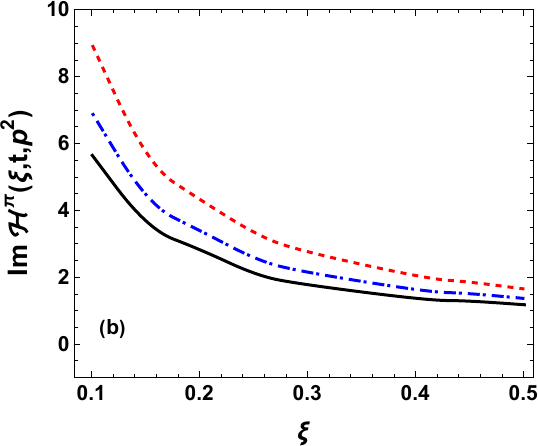} \\
    \includegraphics[width=0.38 \textwidth]{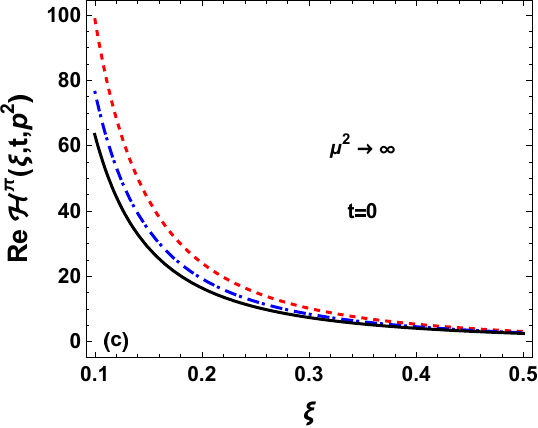}
    \caption{Real and imaginary parts of the Compton form factor of $\pi^+$  in SQM, evolved with LO DGLAP-ERBL equations 
    to $\mu^2=4~{\rm GeV}^2$~(a), and in the asymptotic limit~(b).
    \label{fig:CFFPit1}}
\end{figure}

Another set of important quantities we wish to discuss are the Compton form factors (CFFs) of the pion, which in particular  
enter the cross section for the Sullivan process (see, e.g., \cite{Amrath:2008vx,Diehl:2003ny} and references therein). The DVCS amplitude involves 
the Compton scattering on the partons making up a hadron (cf. Fig~\ref{fig:gpd}). CFFs are obtained
via convolution of the GPDs with a kernel that can be calculated perturbatively in QCD. 
At the leading order, the half-off-shell CFF of the pion (we take $\pi^+$ for definiteness) is equal to
\begin{align}
    \mathcal{H}^\pi(\xi,t,p^2) &= \sum_{p=u,\bar{d}} e_p^2 \int_{-1}^{1} dx H^p(x,\xi,t,p^2,0)\nonumber\\
    & \times\left(\frac{1}{\xi-x -i\epsilon}-\frac{1}{x+\xi -i\epsilon}\right), \label{eq:cdef}
\end{align}
where $p$ indicates a parton and $e_p$ represents its electric charge in units of $e$. Since the perturbative kernel is antisymmetric in $x$, only the $I=0$ parts of the quark 
GPDs from Eq.~(\ref{eq:Hq}), which are also antisymmetric, enter Eq.~(\ref{eq:cdef}). Then
\begin{align}
    {\rm Re}\,\mathcal{H}_\pi(\xi,t,p^2) &= 2\frac{5}{9}  \, \mathbb{P} \int_0^{1} dx \frac{H^0(x,\xi,t,p^2,0)}{\xi-x}, \label{eq:cdefri} \\
    {\rm Im}\,\mathcal{H}_\pi(\xi,t,p^2) &= \frac{5}{9}  \pi H^0(\xi,\xi,t,p^2,0), \nonumber
\end{align}
where $\mathbb{P}$ indicates the principal value integral.

Since the GPDs were evolved with only the LO evolution kernel, we present here the results of only the LO 
CFFs.\footnote{The LO level may not be sufficient, as the NLO effects with the gluonic contributions have been found to be important or dominating in several calculations \cite{Pire:2011st,Moutarde:2013qs}.}
The real and imaginary parts of the LO CFFs are plotted in Fig. \ref{fig:CFFPit1} for $t=0$, and  $ \xi \leq 0.5$. The real part of the CFF displays a significant dependence on the off-shellness of the pion, up to $\sim 65\%$. The deviation from the on-shell limit marginally decreases as the skewness increases. Further, the real part of the CFF is large and positive for $\xi\lesssim 0.25$ and exhibits a smooth decreasing behavior as skewness increases. For $\xi\gtrsim 0.25$ the real part of CFF becomes small and negative.\par
The imaginary part exhibits a somewhat smaller dependence on the off-shellness ($\sim 33\%$), which 
decreases with the increase in $\xi$. The imaginary part is a monotonically decreasing function of the
skewness and is positive for all the values of $\xi$ plotted in Fig. \ref{fig:CFFPit1}.

In the asymptotic limit of $\mu^2\to\infty$, if follows from Eqs.~(\ref{eq:asev}) that the imaginary part of the CFF vanishes, 
\begin{align}
  \lim_{\mu \to \infty}  {\rm Im}\,\mathcal{H}^\pi(\xi,t,p_i^2,p_f^2) &=0,
\end{align}
while the $\xi$-dependence of the real part is given by
\begin{align}
  \lim_{\mu \to \infty}  {\rm Re}\,\mathcal{H}^\pi(\xi,t,p_i^2,p_f^2) &= \frac{25 N_F N^0(\xi,t,p_i^2,p_f^2)}{9 (4 C_F +N_F)} \frac{1}{\xi^2},
\end{align}
which is singular at $\xi \to 0$. Note that at low values of $\xi$, the factor $N^0(t,\xi,p_i^2,p_f^2)$ is dominated by the gravitational form factor $\theta_2(t,p_i^2,p_f^2)$.

\section{Off-shellness of the pion propagator \label{sec:offpi}}

\subsection{General considerations}

\begin{figure*}[t]
    \centering
    \includegraphics[width=0.35\textwidth]{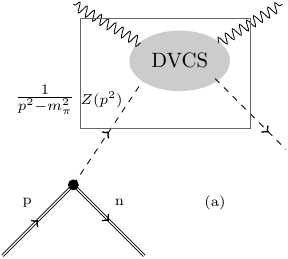} \hspace{1cm} \includegraphics[width=0.35\textwidth]{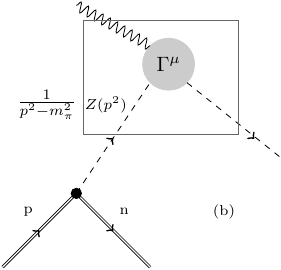}
    \caption{The DVCS (a) and the pion electroproduction (b)~amplitudes, with the off-shell effects in the pion propagator present in $Z(p^2)$. This factor may be absorbed in redefined DVCS or $\Gamma^\mu$ amplitudes, as indicated by the surrounding boxes. 
    Then the pion pole term $1/(p^2-m_\pi^2)$ and not the full pion propagator is used for the exchanged pion. \label{fig:Z}}
\end{figure*}

Whereas in model and phenomenological studies one encounters the problem of off-shellness, the issue is quite subtle.
In the 1990's, within the context
of a possible experimental program to determine off-shell effects in
hadronic form factors, it was realized that the off-shell
effects cannot be measured as a physical observable even at the lowest
orders in the chiral perturbation theory for the case of the pion (see,
e.g., \cite{Fearing:1999im,Scherer:2000hh} and references therein). They are model or scheme dependent,
in particular, they depend on the chosen parametrization of the pion
field. 
If, ideally, one were able to evaluate the full cross section $ep \to en\pi^+$ in a model (or simulate it on the lattice), one could then compare it directly
to the experiment. There, the pion would not be approximated with a pole term or a model propagator, 
but all the hadronic (quark and gluon) processes would contribute to the process,
whereby the off-shell effects would never appear. This utopia, however,
is not only currently impossible, but also not desired, as
theoretically we wish to have components (building blocks) of the
amplitude, such as DVCS, which upon factorization enter various physical processes. So one is bound to an evaluation
of the building blocks, where we apply intermediate
hadronic states, and the off-shellness does need to be tackled with~\cite{Ekstein:1960xkd}.

Up to now we have considered the off-shell effects in GPDs, or the resulting FCCs and generalized form factors. Now, we turn to the 
off-shellness of the pion propagator. 
In general, the pion form factor can be written as a product of the pole term and the pion wave function renormalization, 
\begin{eqnarray}
\Delta(p^2) = \frac{1}{p^2-m_\pi^2} Z(p^2), \label{eq:poleZ}
\end{eqnarray}
where $Z(m_\pi^2)=1$. It is clear that when one considers off-shell effects in a hadronic diagram, e.g., in the Sullivan or electroproduction amplitudes 
of Fig.~\ref{fig:Z}, one needs to 
account for their presence in all the components of the diagram. 
Since it is customary to use in such diagrams the pion pole term as the pion propagator, $1/(p^2-m_\pi^2)$, the remaining $Z(p^2)$ factor in Eq.~(\ref{eq:poleZ}) should be included in the 
amplitude connected to the pion pole. This is explained pictorially in Fig.~\ref{fig:Z}. Therefore, with this arrangement in mind, we should multiply the half-off-shell 
GPDs and the corresponding form factors by $Z(p^2)$.

The simplest example is for the pion electroproduction from Fig.~\ref{fig:Z}(b). The contraction with the leptonic tensor removes the longitudinal 
part from vertex (Eq.~(\ref{eq:genEV})), so for the evaluation of the electroproduction cross section one is left with the part containing $2P^\mu F(t,p^2,0)$ only. 
From WTI (Eq.~(\ref{eq:piF})) for the half-off-shell case we find $Z(p^2)=1/ F(t,p^2,0)$, therefore the vertex incorporating the pion renormalization is 
\begin{eqnarray} 
 V^\mu = 2P^\mu \frac{F(t,p^2,0)}{F(0,p^2,0)}. \label{eq:V}
\end{eqnarray} 
In the situation when the $t$ and $p^2$ dependence is strictly factorized, e.g. in SQM in the chiral limit (cf. Eq.~(\ref{eq:ffvecF})), the offshell dependence in 
$V^\mu$ cancels out exactly and we are left with $V^\mu = 2P^\mu F_V(t)$. In a general case the exact factorization need not occur, 
for instance chiral corrections break it weakly, hence we expect some remnant off-shell effects in $V^\mu$. However, as a starting point, we expect the 
off-shell effects in the pion electroproduction to be small. 

Another comment here concerns the pion-nucleon form factor $G_{\pi NN}$ entering diagrams of Fig.~\ref{fig:Z}. In phenomenological approaches one uses simple 
parametrizations, for instance 
\begin{eqnarray}
G_{\pi NN}(p^2)=G_{\pi NN}(m_\pi^2)\frac{\Lambda_\pi^2 - m_\pi^2}{\Lambda_\pi^2 - p^2}. 
\end{eqnarray}
Ideally, the off-shellness effects in $G_{\pi NN}$ should be computed in the same framework as for the other building blocks of the process, but this would require a uniform and efficient model for both the pion and the nucleon, which we do not have at hand. Then, it is difficult to separate the possible off-shell effects in $V^\mu$ and $G_{\pi NN}$.
Moreover, contributions of other states (for instance, excited pions) also contribute to the hadronic process and again, their contribution is intertwined with the possible 
off-shell effects. 

\subsection{Form factors amended with $Z(p^2)$}

\begin{figure}[t]
    \centering
    \includegraphics[width=0.4 \textwidth]{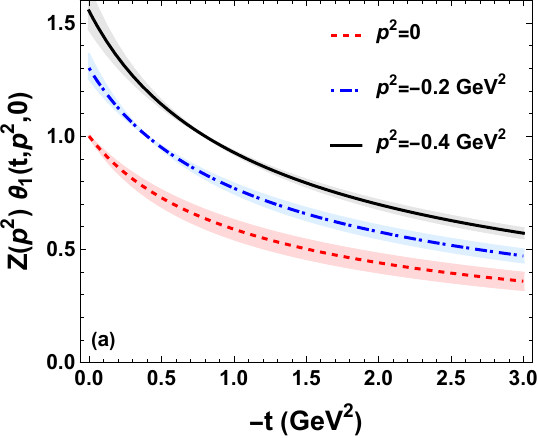}\\\includegraphics[width=0.4 \textwidth]{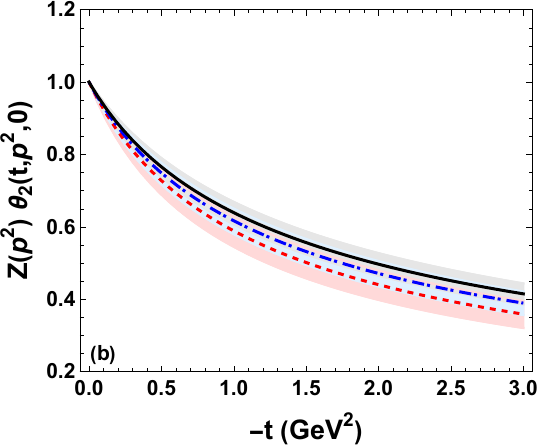}\\\includegraphics[width=0.4 \textwidth]{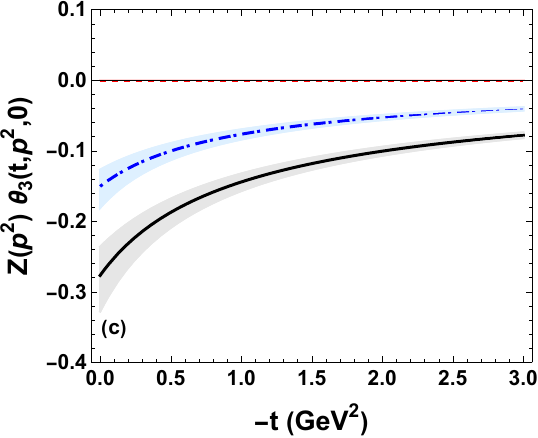}
    \caption{Gravitational form factors $Z(p^2)\theta_1$, $Z(p^2)\theta_2$, and $Z(p^2)\theta_3$ in SQM in the chiral limit.}
    \label{fig:FFG12Full}
\end{figure}

In this subsection we present the half-off shell form factors in SQM in the chiral limit, amended with the pion wave function renormalization factor $Z(p^2)$.
From Eq.~(\ref{eq:invpiprop}-\ref{eq:ffvecG}) we find immediately that 
\begin{eqnarray}
&&Z(p^2)F(t,p^2,0)=\frac{M_V^2}{M_V^2-t}, \\
&& Z(p^2)G(t,p^2,0)=\frac{p^2}{M_V^2-t}
\end{eqnarray}
hence, as already argued in the general discussion above, all dependence of the off-shellness $p^2$ disappears from $F$, while $G$ is strictly 
proportional to $p^2$. 
Correspondingly, for the moment from Eq.~(\ref{eq:normvec2}) we find
\begin{eqnarray}
Z(p^2)N^1(t,\xi,p^2,0)=\frac{M_V^2 - \xi p^2}{M_V^2-t}. \label{eq:z1}
\end{eqnarray}

For the case of the half-off-shell gravitational form factors, where no factorization of $t$ and $p^2$ occurs, we do not find exact cancellation. The results for the 
form factors $Z(p^2)\theta_i$ are presented in Fig.~\ref{fig:FFG12Full}. We note that $Z(p^2) \theta_2$ depends on $p^2$ very weakly. The changes in $Z(p^2) \theta_1$ and 
$Z(p^2) \theta_3$ from their on-shell $t$-dependence are approximately proportional to $p^2$. 

The moment of Eq.~(\ref{eq:normgrav2}) takes a simple exact form at $t=0$, namely 
\begin{eqnarray}
Z(p^2) N_0(0,\xi,p^2,0) = 1 - \xi \frac{\theta_3(0,p^2,0)}{\theta_2(0,p^2,0)} -\xi^2  \frac{\theta_1(0,p^2,0)}{\theta_2(0,p^2,0)}, \nonumber \\
\!\!\hspace*{-6mm} 
\end{eqnarray}
where we have used the general relation from WTI, Eq.~(\ref{eq:theta3}) and Eq.~(\ref{eq:theta20}). Expansion for low $\xi$ and $p^2$ yields 
\begin{eqnarray}
&& Z(p^2)N_0(t,\xi,p^2,0)= 1-\frac{\xi p^2}{M_V^2} + \ldots\text{ .} \label{eq:z2}
\end{eqnarray}

\subsection{GPDs amended with $Z(p^2)$}

\begin{figure}[t]
    \centering
    \includegraphics[width=0.38\textwidth]{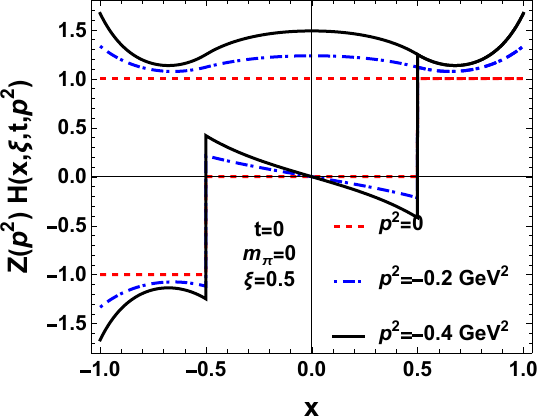}
    \caption{Half-off-shell GPDs of the pion amended with $Z(p^2)$ in SQM at the quark model scale for $t=0$.}
    \label{fig:GPDinFull}
\end{figure}

\begin{figure}[t]
    \centering
    \includegraphics[width=0.4\textwidth]{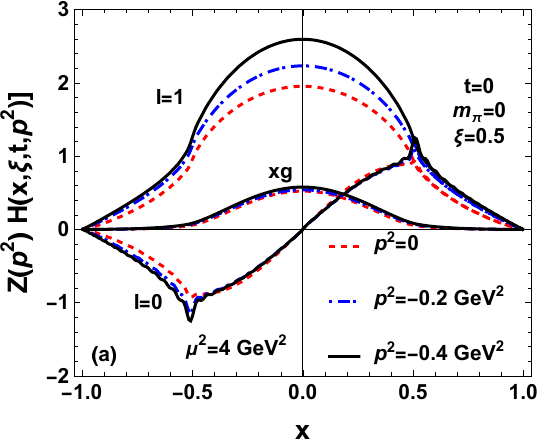}\\\includegraphics[width=0.4\textwidth]{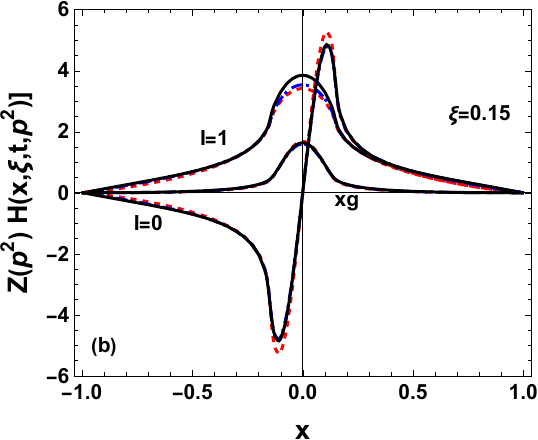}
    \caption{Same as in Fig.~\ref{fig:GPDinFull}, but evolved to $\mu^2=4\text{GeV}^2$ with LO DGLAP-ERBL equations. 
    Panel (a) corresponds $\xi=0.5$ and panel (b) to $\xi=0.15$. 
    \label{fig:GPDevFull}}
\end{figure}

In Fig.~\ref{fig:GPDinFull} we plot the half-off shell GPDs multiplied with the pion wave function renormalization $Z(p^2)$ from SQM in the chiral limit 
at the quark model scale. We take $t=0$ and $\xi=0.5$. 
The corresponding evolved quantities are presented in Fig.~\ref{fig:GPDevFull}.
We note that the normalization is given by the factors $Z(p^2)N^{1,0}$ of Eqs.~(\ref{eq:z1},\ref{eq:z2}). 
Comparing to the curves in Figs.~\ref{fig:GPDin} and \ref{fig:GPDev}, which were normalized to  $N^{1,0}$, we note
an inverted sequence of curves for the corresponding values of $p^2$. In particular, in the present case the normalization decreases (for the assumed positive $\xi$) with increasing $-p^2$, while in Figs.~\ref{fig:GPDin} and \ref{fig:GPDev} it was increasing. We also note by comparing panels (a) and (b) of Fig.~\ref{fig:GPDevFull}
that the effect gets weaker as $\xi$ decreases, in agreement with Eq.~(\ref{eq:z2}).

\subsection{Compton form factors amended with $Z(p^2)$}

\begin{figure}[t]
    \centering
    \includegraphics[width=0.4 \textwidth]{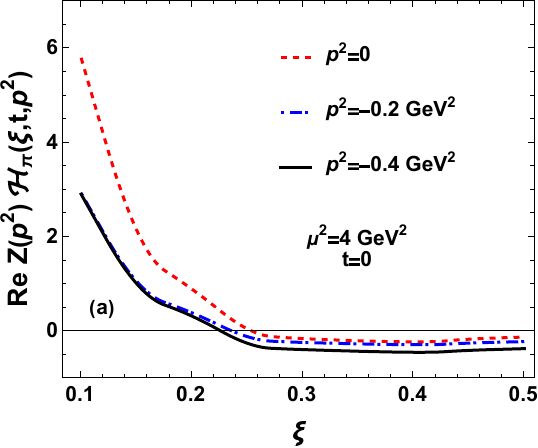}\\\includegraphics[width=0.38 \textwidth]{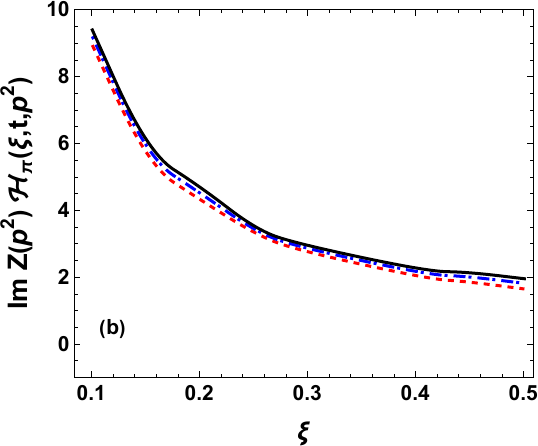}\\\includegraphics[width=0.38 \textwidth]{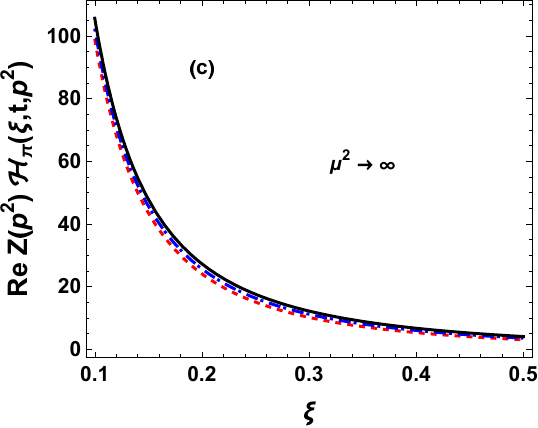}
    \caption{Compton form factors amended with the pion wave function renormalization, $Z(p^2) \mathcal{H}_\pi$, 
    evolved to $\mu^2=4\text{ GeV}^2$ and in the asymptotic limit.}
    \label{fig:CFFFull}
\end{figure}

The features of the Compton form factors multiplied with $Z(p^2)$ reflect the behavior of $Z(p^2) H^0$ discussed in the previous subsection. These quantities are 
plotted in Fig.~\ref{fig:CFFFull}. In particular, we note that the imaginary parts of $Z(p^2)\mathcal{H}^\pi(\xi,t,p^2)$ exhibit a very weak dependence on the off-shellness $p^2$, 
in contrast to Fig.~\ref{fig:CFFPit1}.\par %

In the asymptotic limit, the real part of the CFF is proportional to the GFF $\theta_2(t,p_i^2,p_f^2)$. At $t=0$, the off-shellness of $\theta_2$ is canceled exactly by $Z(p^2)$ (Eqs.~\ref{eq:theta20}, \ref{eq:invpiprop}). The residual off-shell effects arising from $\theta_1$ and $\theta_3$ add up destructively. From Eq.~(\ref{eq:normgrav2}), we see that 
\begin{align}
    \text{Re} Z(p^2)\mathcal{H}_\pi(0,\xi,p^2) &\propto \frac{1-\xi^2}{\xi^2} + \frac{(\xi-1)}{\xi}\frac{p^2}{M_V^2} + \mathcal{O}\left(\frac{p^4}{M_V^4}\right).
\end{align}
With the real part of the CFF dominated by the $1/\xi^2$ behavior, the off-shellness appears only as a negligibly small correction. The imaginary part of the CFF vanishes in the asymptotic limit. \par

The results shown above suggest a strong cancellation of the off-shell effects in CFFs between the GPDs and the pion propagator.
The cross-section for the electroproduction of the pion results from an interference of the DVCS and the Bethe-Heitler amplitudes, with the latter dominating~\cite{Amrath:2008vx}. Hence, any effect of the off-shellness on the DVCS amplitude carries over linearly to the cross section.
Therefore, one should expect only small effects of off-shellness in the electroproduction processes. Since in our model the off-shellness of the EM form factor is largely canceled by $Z(p^2)$, the corrections to the cross-section can arise only from the virtual Compton scattering (VCS) terms. Assuming that these corrections are dominated by the real part of the CFF, we get,
\begin{align}
    \delta\sigma_\text{Tot} &= \Delta_R\frac{2\sigma_\text{VCS} + \sigma_\text{INT}}{\sigma_\text{Tot}}
\end{align}
where the $\sigma$'s are the integrated cross-sections and $\Delta_R=\delta\mathcal{H}_\pi/\mathcal{H}_\pi$. Using the values listed in Table~1 of~\cite{Amrath:2008vx}, we find that $\delta\sigma_\text{Tot}\sim 0.1\Delta_R$. Thus, we expect the corrections to the integrated cross-section to be of the order of a few percent. Since $\Delta_R$ depends on the value of $\xi$, the correction to the differential cross section varies with $\xi$.

\section{Conclusion}\label{sec:SnC}

In this paper we have discussed three groups of topics related to the off-shellness effects in the generalized parton distributions of the pion. 

On the general and formal side, we have demonstrated that in the
absence of the crossing symmetry, the moments of the GPDs pick up odd
powers of the skewness parameter, which results in the appearance of
new form factors that vanish when the pion becomes on-shell, but
otherwise are present.  Under the assumption of the PCAC relations, we
have shown that the Ward-Takahashi identities relate these new
off-shell form factors to the ones that are present for an on-shell
case, as well as to the pion propagator. The electromagnetic and
gravitational form factors pick up a dependence on the off-shellness.

In the second part we have illustrated the formal features of the off-shell
GPDs and the resulting form factors using the Spectral Quark Model
which implements chiral symmetry and incorporates the vector meson dominance principle for the
electromagnetic form factor. The model GPDs exhibit a significant
dependence on the momentum-square of the off-shell pion. Specifically,
the magnitude of the GPDs reduce as the off-shellness increases in
magnitude. The GPDs were then evolved from the quark model scale to
$\mu^2=4~\text{GeV}^2$ using the LO-DGLAP-ERBL evolution equations,
showing that the dependence on the off-shellness holds. This
significant dependence of GPDs carries over to the electromagnetic,
gravitational, and Compton form factors.

Finally, we discuss the effects of the off-shellness in the pion propagator, in conjunction with the half-off-shell GPDs encountered in electroproduction processes. 
With the Ward-Takahashi identities and the derived model formulas we have shown that the combined off-shell effect in the electromagnetic form factor or the Compton form factor at low skewness are tiny. Therefore the combined effects of off-shellness in the pion electroproduction processes is expected to be very small, at $\lesssim 5\%$. This means that naive estimates, not taking into account off-shellness, can be numerically correct, 
but for a nontrivial reason stemming from general considerations involving the Ward-Takahashi identities for the electromagnetic and gravitational vertices.

\section*{Acknowledgements}
We are grateful to Krzysztof Golec-Biernat for providing the QCD evolution code and to the authors of Ref.~\cite{Hackett:2023nkr} for the data used in Fig.~\ref{fig:ffcomplatt}. 
VS acknowledges the support by the Polish National Science Centre (NCN), grant 2019/33/B/ST2/00613, WB by the Polish National Science Centre (NCN), grant 2018/31/B/ST2/01022, and ERA by project PID2020-114767GB-I00 funded by MCIN/AEI/10.13039/501100011033 as well as Junta de Andaluc{\'i}a (grant FQM-225).

\appendix

\section{Derivation of WTIs \label{app:WTIder}}

In this Appendix, we present and discuss the standard derivations of the WTIs for off-shell pions. 
We follow the notation and conventions of~\cite{Rudy:1994qb}, in particular a positively charged pion enters the vertex with momentum $p_i$, and leaves 
with momentum $p_f$.
Consider the full (unamputated) vertex $G^{\mu\mu_2\cdots\mu_n}(p_i,p_f)$ 
representing a matrix element of a local operator $\mathcal{O}^{\mu\mu_2\cdots\mu_n}$. By definition,

\begin{eqnarray}
&&\hspace*{-2mm}  (2\pi)^4\delta^{(4)}(p_f-p_i-q) G^{\mu\mu_2\cdots\mu_n}(p_i,p_f) = \int d^4x \,d^4y \,d^4z  \nonumber \\ 
&& \hspace*{-2mm} \times e^{i(p_f\cdot x - p_i\cdot y - q\cdot z)} \langle 0|T( \phi^+(x)\phi^-(y) \mathcal{O}^{\mu\mu_2\cdots\mu_n}(z))|0\rangle. \label{eq:Gdef}
\end{eqnarray}    
Upon contraction with $q_\mu$ and partial integration one gets
\begin{eqnarray}
&&\hspace*{-1mm}  (2\pi)^4\delta^{(4)}(p_f\!-\!p_i\!-\!q)q_\mu G^{\mu\mu_2\cdots\mu_n}(p_i,p_f) \!= \!- \!i \! \int \!\! d^4x \, d^4y \,d^4z \nonumber \\
&&\hspace*{-1mm}  \times e^{i(p_f\cdot x - p_i\cdot y - q\cdot z)} \frac{\partial}{\partial z^\mu}\langle 0|T( \phi^+(x)\phi^-(y) 
\mathcal{O}^{\mu\mu_2\cdots\mu_n}(z)|0\rangle. \nonumber \\ \label{eq:preWTIG}
\end{eqnarray}
For conserved currents, $\partial/\partial z^\mu\, \mathcal{O}^{\mu\mu_2\cdots\mu_n}=0$, one finds that
\begin{eqnarray}
&&\frac{\partial}{\partial z^\mu}\langle 0|T( \phi^+(x)\phi^-(y) \mathcal{O}^{\mu\mu_2\cdots\mu_n}(z)|0\rangle = \\
&& ~~~~~~ \delta(x^0-z^0) \langle 0|T([\mathcal{O}^{0\mu_2\cdots\mu_n}(z), \phi^+(x)]\phi^-(y) |0\rangle \nonumber\\ 
&& ~~ + \delta(y^0-z^0) \langle 0|T(\phi^+(x)[\mathcal{O}^{0\mu_2\cdots\mu_n}(z), \phi^-(y)] |0\rangle. \nonumber
\end{eqnarray}

The WTI for the given vertex is obtained by imposing the appropriate equal time commutation relations.
For the electromagnetic current one uses
\begin{eqnarray}
\delta(z^0-u^0)[J^0(z),\phi^\pm(u)] = \pm\delta^{(4)}(z-u) \phi^\pm(z). \label{eq:EMcom}
\end{eqnarray}
Thus, Eq.~(\ref{eq:preWTIG}) becomes
\begin{eqnarray}
&&    (2\pi)^4\delta^{(4)}(p_f-p_i-q)q_\mu G^{\mu}(p_i,p_f) = i\int d^4x \,d^4y \times \nonumber \\ 
&&   ~\left( e^{i (p_f - p_i-q) \cdot y+i p_f \cdot (x-y)} - e^{i (p_f - p_i-q) \cdot x+i p_i \cdot (x-y)} \right ) \times \nonumber \\
&& ~~~~~~~~~~~\langle 0|T( \phi^+(x)\phi^-(y)|0\rangle, 
\end{eqnarray}
which yields the relation,
\begin{eqnarray}
q_\mu G^\mu(p_i,p_f) = \Delta(p_f^2)-\Delta(p_i^2), \label{eq:preWTIEM}
\end{eqnarray}
where 
\begin{eqnarray}
\Delta(p^2) = \int d^4u \, e^{-i p \cdot u}\langle 0|T( \phi^+(0)\phi^-(u)|0\rangle
\end{eqnarray}
is the pion propagator.

The derivation of WTI for the stress-energy tensor proceeds along similar lines, but is more involved, since the commutation relations contain the derivatives
with respect to time. The separation of the time derivatives in the operator and the time-ordering has been know to be subtle. It
requires the use of the $T^\ast$ products, where the time differentiation is pulled out in front of the time ordering. With the commutation relation
for the energy-stress tensor with one time component,
\begin{eqnarray}
\hspace{-6mm} \delta(z^0-u^0)[\Theta^{\mu0}(z),\phi^a(u)] &= -i\delta^{(4)}(z-u)\frac{\partial}{\partial z_\mu}\phi^a(z), \label{eq:ESTcom} \nonumber \\
\end{eqnarray}
where, $a$ represents the isospin index, we get,
\begin{eqnarray}
&& \frac{\partial}{\partial z^\mu}\langle 0|T^\ast( \phi^a(x)\phi^b(y) \Theta^{\mu\nu}(z)|0\rangle = \\
&& -i\delta^{(4)}(x-z) \langle 0|T^\ast\left(\frac{\partial}{\partial x_\nu} \phi^a(x)\phi^b(y)\right) |0\rangle \nonumber\\
&&    -i \delta^{(4)}(y-z) \langle 0|T^\ast\left(\phi^a(x)\frac{\partial}{\partial y_\nu}\phi^b(y)\right) |0\rangle. \nonumber
\end{eqnarray}    

Repeating the steps leading to the electromagnetic WTI, one gets the relation,
\begin{eqnarray}
q_\mu G^{\mu\nu}(p_i,p_f) &= p_i^\nu \Delta(p_i^2) - p_f^\nu\Delta(p_f^2).  \label{eq:preWTIGFF}
\end{eqnarray}
This relation was first derived by Brout and Englert~\cite{Brout:1966oea} from just the general gravitational covariance.

Some remarks and discussion are in place. The above derivations were carried out with a tacit assumption that the pion is an elementary field
satisfying canonical commutation relations, which allows 
for disregarding possible Schwinger terms in the commutation relations (\ref{eq:EMcom}) and  (\ref{eq:ESTcom}). 
Note that for the case of charge algebra, i.e. when we integrate  (\ref{eq:EMcom}) over $d^3z$, we find
\begin{eqnarray}
[I_3,\phi^\pm(u)]_{\rm ET} = \pm \phi^\pm(z), \label{eq:EMcom2}
\end{eqnarray}
where $I_3$ is the third component of the isospin operator. Similarly, from (\ref{eq:ESTcom}) it follows that 
\begin{eqnarray}
[P^{\mu},\phi^a(u)]_{\rm ET} = -i \frac{\partial}{\partial z_\mu}\phi^a(z), \label{eq:ESTcom2}, 
\end{eqnarray}
where $P^\mu$ is the four-momentum operator which is a generator of translations. 

In the current-algebraic derivations one may depart from the assumption of the elementary nature of the pion, but one tacitly assumes that the pion 
is the interpolating field satisfying the strong PCAC relation of the form
\begin{eqnarray}
\phi^a(u)=\partial^\mu A^a_\mu(u), \label{eq:pcac}
\end{eqnarray}
where $A^a_\mu$ is the axial vector current. This assumption is at the core of the derivations in~\cite{Schnitzer:1967zzb,Naus:1989em,Rudy:1994qb} for
the electromagnetic case, or in~\cite{Raman:1971jg} for the stress-energy tensor case.
We note that in our model the pion is not 
elementary, as it is a composite quark-antiquark object, but it does satisfy PCAC of Eq.~(\ref{eq:pcac}). Thus, it naturally complies to the WTIs of Eqs.~(\ref{eq:preWTIEM},\ref{eq:preWTIGFF}).
This is exemplified with the explicit forms obtained in SQM, such as Eqs.~(\ref{eq:invpiprop}-\ref{eq:ffgrav}),
which satisfy all the WTI-based relations given in Sec.~\ref{sec:ff}.

\section{Passarino-Veltman functions \label{app:PaVe}}

With the Klein-Gordon denominators 
\begin{eqnarray}
D_l=l^2-\omega^2+i \epsilon,
\end{eqnarray}
the Passarino-Veltman functions needed for the evaluation of the half-off-shell form factors are defined as 
\begin{eqnarray}
&& i \pi^2 B_0(\omega^2, v^2) = \int\!\! d^4k \frac{1}{D_k D_{k+v}}, \nonumber \\
&&  i \pi^2 C_0(\omega^2, t,p^2) = \int\!\! d^4k \frac{1}{D_k D_{k-p} D_{k+q}},
\end{eqnarray} 
where $\omega$ is the quark mass, and $v=q$ or $p$. We note that the Passarino-Veltman functions are analytic in all their arguments.
Upon the Wick rotation, with the corresponding Euclidean momenta denoted with capital letters, we have the notation
\begin{eqnarray}
&& \pi^2 B_0(\omega^2, v^2) = \int\!\! d^4K \frac{1}{[K^2+\omega^2][(K+V)^2+\omega^2]}, \nonumber \\
&&  i \pi^2 C_0(\omega^2, t,p^2) = - \int\!\! d^4K \times \\
&& ~~~~~\frac{1}{[K^2+\omega^2][(K-P)^2+\omega^2][(K+Q)^2+\omega^2]}, \nonumber
\end{eqnarray}

\section{One-loop expressions for half-off-shell form factors \label{app:1loop}}
 
Here we provide the general one-quark-loop expressions for the half-off-shell electromagnetic and gravitational form factors of the pion. 
Formulas for general $p_i^2$ and $p_f^2$ can also be given, but they are lengthy.
The half-off shell electromagnetic form factors in the chiral limit are given by
\begin{align}
    F(t,p^2,0) &= \frac{N_c}{4f^2} \int_C d\omega \rho(\omega) \omega^2 \frac{tB_0(\omega^2,t)-p^2B_0(\omega^2,p^2)}{t-p^2},\\
    G(t,p^2,0) &= \frac{N_c}{4f^2} \int_C d\omega \rho(\omega) \omega^2 \frac{p^2 [B_0(\omega^2,t) - B_0(\omega^2,p^2)]}{t-p^2},
\end{align}
whereas the gravitational form factors are 
\begin{widetext}
\begin{align}
     \theta_1(t,p^2,0) &=  \frac{N_c}{4\pi^2f^2} \int_C d\omega \rho(\omega) \omega^2 
          \frac{p^4B_0(\omega^2,p^2) + t(t-p^2)B_0(\omega^2,t)+2\omega^2(t-p^2)(t-2p^2)C_0(\omega^2,t,p^2)}{(t-p^2)^2}, \\
    \theta_2(t,p^2,0) &=  \frac{N_c}{4\pi^2f^2} \int_C d\omega \rho(\omega) \omega^2 
         \frac{t^2B_0(\omega^2,t) + p^2(p^2-2t)~B_0(\omega^2,p^2)+2\omega^2t(t-p^2)C_0(\omega^2,t,p^2)}{(t-p^2)^2}, \nonumber\\
    \theta_3(t,p^2,0) &=  \frac{N_c}{4\pi^2f^2} \int_C d\omega \rho(\omega) \omega^2 
         \frac{p^2}{(t-p^2)^2} \left(tB_0(\omega^2,p^2) - tB_0(\omega^2,t) + 2\omega^2(p^2-t)C_0(\omega^2,t,p^2)\right),\nonumber\\
    \theta_4(t,p^2,0) &=  \frac{N_c}{4\pi^2f^2} \int_C d\omega \rho(\omega) \omega^2 
          \frac{p^2 (p^4-p^2 t+t^2)B_0(\omega^2,p^2) -p^4t B_0(\omega^2,t) +2 p^4  (p^2-t) \omega^2 C_0(\omega^2,t,p^2)}{\left(p^2-t\right)^2},\nonumber
\end{align}    
\end{widetext}
where $B_0$ and $C_0$ are the Passarino-Veltman two-point integrals from Appendix~\ref{app:PaVe}. 

The square of the pion decay constant is~\cite{Broniowski:2007si}
\begin{eqnarray}
f^2= \frac{N_c}{4\pi^2} \int_C d\omega \rho(\omega) \omega^2 B_0(\omega^2,0),
\end{eqnarray}
with which one can verify the proper limits of $F$ and $\theta_2$ at $t=p^2=0$.
In SQM,
\begin{eqnarray}
f^2=\frac{N_c M_V^2}{24 \pi^2}. \label{eq:f}
\end{eqnarray}

We note that for the 
evaluation in SQM we need the spectral moments
\begin{eqnarray}
&&   \frac{N_c}{4\pi^2f^2}\int_C d\omega \rho(\omega) \omega^2 B_0(\omega^2,u^2) = \frac{M_V^2}{M_V^2-u^2},\\
&&   \frac{N_c}{4\pi^2f^2} \int_C d\omega \rho(\omega) \omega^4 C_0(\omega^2, t,p^2) = \nonumber \\
&& ~~\frac{M_V^4}{2 (t-p^2)}\left(\frac{M_V^2}{M_V^2-p^2 }+\log\left[\frac{p^2 -M_V^2}{t-M_V^2}\right]+\frac{M_V^2}{t-M_V^2}\right). \nonumber
\end{eqnarray}
The form factors reduce to the ones given by, Eq.~(\ref{eq:ffvecF}) and Eq.~(\ref{eq:ffvecG}). It is 
straightforward to show that the form factors follow the relation given in Eq.~(\ref{eq:piG}). 
When evaluated in SQM, the above expressions reduce to Eq.~(\ref{eq:ffgrav}).

\section{One-loop functions for the GPDs \label{app:IJ}}

The formulas collected in his Appendix follow straightforwardly from the derivation in Appendix~B of~\cite{Broniowski:2007si}. Here we use the symmetric convention.

The two-point functions needed for the evaluation of GPDs are
\begin{eqnarray}
&& I[x,\mp \xi,p_{i,f}^2]\equiv -i \frac{4 N_c \omega^2}{f^2} \int \frac{d^4k}{(2\pi)^4} \frac{\delta(k \cdot n-x)}{D_{k-P} D_{k\pm q/2}}
\nonumber \\
&&~~=\Theta[(1-x)(x\pm\xi)]\frac{N_c \omega^2}{4\pi^2 f^2(1\pm\xi)} \times \nonumber \\
&&~~~~~ \int_0^\infty du \frac{1}{u+\omega^2 - \frac{(1-x)(x\pm\xi))}{(1\pm\xi)^2} p_{i,f}^2},
\end{eqnarray}
In SQM, the evaluation yields~\cite{Broniowski:2007si}
\begin{eqnarray}
I_{\rm SQM}[x,\mp \xi,p_{i,f}^2])= \frac{\Theta[(1-x)(x\pm\xi)]}{(1\pm\xi)\left( 1- \frac{(1-x)(x\pm\xi)}{(1\pm\xi)^2} \frac{4p_{i,f}^2}{M_V^2} \right )^{5/2}}. \nonumber \\
\end{eqnarray}
where relation (\ref{eq:f}) has been used. 

The needed three-point function can be written in the form
\begin{eqnarray}
&& J(x,\xi,t,p_i^2,p_f^2)\equiv \\ && i \frac{4 N_c \omega^2}{f^2} \int \frac{d^4k}{(2\pi)^4} \frac{\delta(k \cdot n-x)}{D_{k-P} D_{k+q/2}D_{k- q/2}} =
\nonumber \\
&&\int_0^1 \!\!\!\!dy  \int_0^1 \!\!\!\!dz \, \Theta(1\!-\!y\!-\!z) \delta[x\!-\!z\! -\!\xi(1\!-\!2y\!-\!z)]{\cal F}(y,z,t,p^2), \nonumber 
\end{eqnarray}
where the {\em double distribution} is
\begin{eqnarray}
&& {\cal F}(y,z,t,p_i^2,p_f^2) = \\
&&\frac{N_c \omega^2}{4\pi^2 f^2 [\omega^2-y z p_f^2+z p_i^2 (y+z-1)+t y (y+z-1)]}. \nonumber
\end{eqnarray}
In SQM
\begin{eqnarray}
{\cal F}_{\rm SQM}(y,z,t,p^2) = \frac{6}{\left [ 1-\frac{4[y z p_f^2-z p_i^2 (y+z-1)-t y (y+z-1)]}{M_V^2} \right ]^{5/2}}.\nonumber\\
\end{eqnarray}

\bibliography{Ref}

\end{document}